\documentclass[conference]{IEEEtran}
\IEEEoverridecommandlockouts

\pdfoutput=1

\usepackage{amssymb}
\usepackage{bm}
\usepackage{enumitem}
\usepackage{comment}
\usepackage{booktabs} % For formal tables
\usepackage[ruled,vlined,linesnumbered]{algorithm2e}
\usepackage{xcolor}
\usepackage{pgfplots}
\usepackage{tikz}
\usepackage{tikzscale}
\usepackage{array}
\usepackage{booktabs}
\usepackage{array,makecell}
\usepackage{boldline}
\usepackage{placeins}
\usepackage{soul}
\usepackage{textcomp}
\usepackage{filecontents}
\usepackage{subcaption}
\usepackage{hyperref}
\usetikzlibrary{external}
	\usepgfplotslibrary{colorbrewer}
    
\definecolor{bblue}{HTML}{4F81BD}
\definecolor{rred}{HTML}{C0504D}
\definecolor{ggreen}{HTML}{9BBB59}
\definecolor{ppurple}{HTML}{9F4C7C}

\pgfplotsset{
    legend image with text/.style={
        legend image code/.code={%
            \node[anchor=center] at (0.3cm,0cm) {#1};
        }
    },
}

\input{figures/setting}

\begin{document}

\def\HS{\hspace{\fontdimen2\font}}

\title{PanJoin: A Partition-based Adaptive Stream Join}
%\titlenote{Produces the permission block, and
%  copyright information}
%\subtitle{Extended Abstract}
%\subtitlenote{The full version of the author's guide is available as
 % \texttt{acmart.pdf} document}

\author{\IEEEauthorblockN{Fei Pan$^1$ \quad Hans-Arno Jacobsen$^2$  }
\IEEEauthorblockA{Electrical \& Computer Engineering, University of Toronto, Toronto, Canada \\
%Toronto, Canada \\
$^1$ fei.pan@mail.utoronto.ca \quad $^2$jacobsen@eecg.toronto.edu}
%\and
%\IEEEauthorblockN{Hans-Arno Jacobsen}
%\IEEEauthorblockA{ECE, University of Toronto, Canada \\
%Toronto, Canada \\
%jacobsen@eecg.toronto.edu}
}

\maketitle

%\renewcommand{\shortauthors}{F. Pan et al.}
%\newcommand{\subscript}[2]{$#1 _ #2$}

\begin{abstract} 

In stream processing, stream join is one of the critical sources of performance bottlenecks. The sliding-window-based stream join provides a precise result but consumes considerable computational resources. The current solutions lack support for the join predicates on large windows. These algorithms and their hardware accelerators are either limited to equi-join or use a nested loop join to process all the requests. 

In this paper, we present a new algorithm called PanJoin which has high throughput on large windows and supports both equi-join and non-equi-join. PanJoin implements three new data structures to reduce computations during the probing phase of stream join. We also implement the most hardware-friendly data structure, called BI-Sort, on FPGA. Our evaluation shows that PanJoin outperforms several recently proposed stream join methods by more than 1000x, and it also adapts well to highly skewed data.

\end{abstract}

%\keywords{ACM proceedings, \LaTeX, text tagging}

%\settopmatter{printacmref=false}

\pgfplotscreateplotcyclelist{my black white}{%
solid, every mark/.append style={solid, fill=gray}, mark=*\\%
dotted, every mark/.append style={solid, fill=gray}, mark=square*\\%
densely dotted, every mark/.append style={solid, fill=gray}, mark=otimes*\\%
loosely dotted, every mark/.append style={solid, fill=gray}, mark=triangle*\\%
dashed, every mark/.append style={solid, fill=gray},mark=diamond*\\%
loosely dashed, every mark/.append style={solid, fill=gray},mark=*\\%
densely dashed, every mark/.append style={solid, fill=gray},mark=square*\\%
dashdotted, every mark/.append style={solid, fill=gray},mark=otimes*\\%
dashdotdotted, every mark/.append style={solid},mark=star\\%
densely dashdotted,every mark/.append style={solid, fill=gray},mark=diamond*\\%
}

\section{Introduction}

Stream processing techniques are used in many modern applications to analyze data in real time. In stream processing, stream join is a commonly used operator for relating information in different streams. Many researchers \cite{gulisano2016scalejoin, teubner2011soccer, ananthanarayanan2013photon, lin2015scalable} consider stream join to be one of the most critical and expensive operations in stream processing.

The definition of stream join is extended from the relational (theta) join. A commonly used model is sliding-window-based stream join (henceforth referred to as \textit{stream join} in this paper), where each stream maintains a sliding window and each incoming tuple from one stream is joined with the sliding window(s) of other stream(s). Stream join has been intensively studied over the past two decades \cite{kang2003evaluating, xie2007survey, teubner2011soccer, roy2014low, vitorovic2016squall, najafi2016splitjoin, lin2015scalable, gulisano2016scalejoin}. The main challenge in processing a stream join is producing results at runtime with high throughput, particularly when the window size is large and the input data rate is high. Recent works \cite{teubner2011soccer, roy2014low, najafi2016splitjoin, gulisano2016scalejoin} attempted to address this challenge by dividing a stream window into several subwindows and assigning them to multiple processors or join cores that work in parallel. In this way, their solutions can handle several thousand input tuples per second with a window size of several million. The limitation of these works is that they mainly parallelize stream join at the architectural level: there are few discussions on how to parallelize subwindow internally or design a data structure to accelerate the processing. Furthermore, a recent report from LinkedIn \cite{noghabi2017samza} shows a real-world requirement to handle input rates of 3.5K-150K per second, which is higher than the maximum throughput of the aforementioned works.

In this paper, we first propose a new architecture generalized from an algorithm named Red Black indexing tree based Symmetric Nested Loop Join (RBSNLJ) \cite{ya2006indexed}. In the new architecture, the window is divided into subwindows based on the arrival time of tuples, and subwindows are chained similar to a circular buffer. The new tuples are only inserted into the newest subwindow, and the oldest subwindow(s) is(are) expired as a whole, while the middle subwindows remain unchanged. This architecture is suitable for implementing a complex data structure such as a tree inside each subwindow, since expiration is highly simplified and the processing overhead caused by the remaining expired tuples is decreased by the number of subwindows. We then propose three new data structures that are specifically designed for the subwindows of this architecture. Each of the data structures has advantages over the others under different categories of configuration, e.g., different subwindow sizes or selectivity values. The main idea of these data structures is to further divide a subwindow into several partitions based on the tuple value. In this way, we can reduce a considerable amount of memory access for probing operation compared with the nested-loop join used in the related works \cite{teubner2011soccer, najafi2016splitjoin, gulisano2016scalejoin}, thereby significantly accelerating the overall stream join by more than 1000x. The three data structures can also process requests with multiple threads, and we design some mechanisms for our data structures to handle highly skewed data, where the tuple value is not evenly distributed among the value range. Therefore, we name our algorithm PanJoin (Partitioned Adaptive uNiformization join), which partitions and parallelizes a sliding window at both the architectural level and the subwindow level, as well as adaptively manages highly skewed data to achieve performance as good as managing data following a uniform distribution. Furthermore, since FPGA is commercially applicable in several cloud computing service providers (e.g., IBM, Amazon and Microsoft) and becomes increasingly popular, we decide to implement the most hardware-friendly data structure, named BI-Sort, on FPGA to achieve high throughput and high energy efficiency.  

%Our goal is to partition the window evenly and bring adaptiveness to data skew. Figure \ref{fig:overview} illustrates an overview of our algorithm. The entire sliding window of one stream is divided into subwindows. In each subwindow, tuples are range partitioned and managed by several data structures (e.g., partition table). New tuples are inserted into the newest subwindow, and old tuples expire from the oldest subwindow. All data in the other subwindows remain unchanged. For probing, the tuple from the partner stream will join with the corresponding partition(s) in each subwindow. 

%Since heterogeneous computing has gained favor in recent research on database system, in this work, we implement our algorithm on both CPU and FPGA. Specifically, we use High-Level Synthesis (HLS) to implement our hardware version on FPGA. HLS is a technology that can convert software code to hardware. By using HLS, we can share the code between software and hardware solutions in our project. In order to do this, we tailored our algorithm so that it can be both software and hardware friendly. By showing the potential of HLS in this work, we believe HLS will play a much more important rule in high-performance database applications and research in the future.

Our experiment shows that PanJoin can handle an input rate of 10M-28M tuples per second for a window size of 8M-1G on a cluster, which is more than 1000x faster than the recently proposed solutions \cite{teubner2011soccer, roy2014low, najafi2016splitjoin, gulisano2016scalejoin}.  Our data structures adapt well to highly skewed data from the real world. The subwindows on FPGA also provide energy-efficient performance compared to the subwindows on CPU.    

%by using one core of a Core i5 processor, our algorithm can outperform several other stream join solutions by 22-144x. It also achieves a performance as high as 255K tuples per second with a window size of 256M. As far as we know, this is the highest performance that has been achieved among recent works on stream join. Our algorithm is also effective in adapting the skewed data: if the new distribution is a uniform or normal distribution, the adjustment algorithm needs only two iterations to calculate the new values of the partition table.

The contributions of this paper are four-fold:
\begin{enumerate}
\item A new stream join architecture which significantly simplifies expiration operation and avoids communication between worker nodes.
\item Three novel data structures which remarkably reduce the comparisons in probing opeartion over nested-loop join by more than three orders of magnitude.
\item Several innovative data structures (e.g., Linked List Adaptive Table) and algorithms to implement the three new data structures or provide efficient storage strategy.
\item An FPGA solution which has more than 4x energy efficiency over the corresponding CPU solution.
\end{enumerate}

%is a stream join algorithm that has high performance, adapts to highly skewed data, and supports non-equi-join. To implement range partitioning, we also propose an innovative data structure to store the data called \textit{Linked List Adaptive Table}, or LLAT. Our proposed adjustment algorithm can also be used in other applications such as constructing indexes for unknown data. Our algorithm can inspire more ongoing effort to implement high performance or distributed stream processing systems. We prototype our algorithm on different hardware platforms (CPU and FPGA) and compare the performance on each platform. 

%In addition, we also plug PanJoin into Apache Storm to explore the real-time use case of our algorithm. 

%Furthermore, we present the potential of the High-Level Synthesis (HLS) technology in implementing FPGA applications. We believe HLS will play a much more important rule in high-performance database applications and research in the future.

The organization of this paper is as follows: Section \ref{sec:rel} shows and discusses the related works; Section \ref{sec:pan} presents the PanJoin algorithm; Section \ref{sec:imp} presents the implementation of PanJoin on FPGA; Section \ref{sec:eva} shows the performance results of PanJoin.

\section{Related Work} \label{sec:rel}
%There are many works on join algorithms and acceleration of join processing. However, most works \cite{balkesen2013multi, albutiu2012massively, balkesen2015main, halstead2015fpga} focus on non-streaming cases. Generally, it is not straightforward to map these works directly to stream join processing, since they have different mechanisms and requirements: traditional join processes a large batch of data at once, while stream join has to produce results on the fly. Therefore, we mainly consider stream join processing and its hardware acceleration. % as well as hardware accelerators for relational database processing.

In this section, we mainly consider stream join and its hardware acceleration. We do not include non-streaming cases because it is not straightforward to map these works directly to stream join processing: traditional join processes a large batch of data at once, while stream join has to produce results on the fly. 
%\subsection{Stream Join Algorithms}
\textit{Symmetric Hash Join} (SHJ) \cite{wilschut1993dataflow} is one of the first proposals for bounded input stream-like data. It continuously updates its two hash tables (one for each relation) that stores the incoming tuples while it receives tuples of each relation in turn. 
%Every new tuple is hashed twice: once in the local relation to find its bucket for storage, and once in the partner relation to find the bucket it should probe. 
SHJ uses hash functions to reduce the comparison for equi-join. 
It also has an extension called XJoin \cite{urhan2001dynamic} to handle disk-resident data. 
XJoin becomes a basis for many later designed algorithms, such as \textit{Rate-based Progressive Join} (RPJ) \cite{tao2005rpj} which uses a smarter memory replacement algorithm. 
After XJoin, 
Kang et al. presented the first formalized 3-step paradigm for window-based stream join \cite{kang2003evaluating}.   

%Though SHJ and the XJoin family are designed for relational database, their algorithm is easily mapped to stream processing cases.

%ripple join, progressive merge join

Red Black indexing tree based Symmetric Nested Loop Join (RBSNLJ) \cite{ya2006indexed} is one of the first works on processing non-equi stream joins. RBSNLJ partitions the sliding window into subwindows based on the arrival time of tuples. The complex data structures (e.g., the red-black trees in RBSNLJ) are built in subwindows. Only the data structure in the newest subwindow is updated at runtime and the other subwindows reuse their data structures for probing operation. 
%A similar concept can be found in some later works \cite{gedik2009celljoin}. 
In this way, the cost of updating the complex data structure is significantly reduced, which increases the overall performance. However, RBSNLJ does not explore how to process a subwindow in parallel. The red-black tree is also unnecessarily complicated and generates a large number of random memory accesses during processing, compared with the data structures used in PanJoin. 

To reduce comparisons during probing, some research works focus on increasing the parallelism of join processing. CellJoin \cite{gedik2009celljoin} divides a sliding window into subwindows based on the arrival time of tuples, which is similar to RBSNLJ. CellJoin maps subwindows to processing cores on heterogeneous multicore Cell processors. Because CellJoin implements nested-loop-based join processing, its performance is limited to several thousand tuples per second on a 15-minutes window (several million tuples in total). Nevertheless, we believe CellJoin can easily be extended to support our PanJoin by adding our specially designed data structures into subwindows. 

(Low Latency) Handshake Join \cite{teubner2011soccer, roy2014low} parallelizes the join processing by maintaining a bidirectional data flow similar to a doubly linked list. Each node (subwindow) in the flow is mapped to a processing core. New tuples flow from the starting node to the ending node and join with the tuples of the opposite stream saved in all bypassed node. The performance of handshake join is also constrained by its nested-loop-based join processing.    

SplitJoin \cite{najafi2016splitjoin} parallelizes join processing in a more straightforward manner. Rather than forwarding tuples through the data flow, SplitJoin stores each tuple in a fixed node. It also splits storage and probing into two separate steps, where probing can be further divided into several independent processes that run in parallel. BiStream \cite{lin2015scalable} processes stream join in a similar fashion and is applied on large-scale distributed stream join systems. BiStream also implements indexes to accelerate equi-join and non-equi-join. Another similar work called ScaleJoin first merges all incoming tuples into one stream, and then it distributes them to \textit{processing threads} (PTs) \cite{gulisano2016scalejoin}. Each tuple is dispatched to all PTs but is stored in only one PT in a round robin fashion. However, (Low Latency) Handshake Join, SplitJoin and ScaleJoin lack discussions on the data management inside the subwindows (nodes or PTs) while PanJoin provides three new data structures for users to chose. Their subwindows also update and expire the tuples frequently, which is inefficient for a complex data structure to manage. PanJoin solves this problem by providing an architecture with a highly simplified expiration mechanism.

%In Section \ref{sec:comp}, we show that our PanJoin outperforms SplitJoin and ScaleJoin using only one core on a two-core laptop CPU.

%\subsection{Hardware Acceleration on Stream Join}

Recently FPGAs have received increasingly more attention in accelerating stream processing. Pioneering works show the tremendous potential of FPGAs in stream processing \cite{hagiescu2009computing, woods2010complex, sadoghi2012multi}. 
Handshake Join has an FPGA implementation \cite{teubner2011soccer}. ScaleJoin \cite{gulisano2016scalejoin} also has an FPGA version \cite{kritikakis2016fpga} that uses 4 Virtex-6 FPGAs. % With only 30\% of resource cost on FPGAs, this implementation outperforms the original ScaleJoin on a 48-core system by a factor of 19x in processing rate and 4x in throughput.
In addition to a pure stream join processor, M. Najafi et al. suggest a more general FPGA-based stream processing architecture, Flexible Query Processor (FQP) \cite{najafi2015configurable}, which perform stream join similar to Handshake Join. FQP introduces OP-Block, a connectable stream processing unit, for constructing a Flexible Query Processor to process complex queries. Each OP-Block can be configured as a join core, a projection filter or a selection filter. FQP maps stream processing queries to a chain of OP-Blocks, thus handshake join can be realized as a chain of OP-Blocks which are configured as join cores. Similar to handshake join, in FQP each tuple is passed from the first partition stored in the first OP-Block (near input) to the next partition (OP-Block). Therefore FQP also dynamically partitions the window based on the arrival time of each tuple. To collects result, each OP-Blocks is connected to a result filter which inserts the result into a result aggregation buffer, and then the result is output to the result stream. FQP implements nested-loop join inside each OP-Block. Therefore, as a pioneer work, the peak throughput of FQP is limited. Here, both FQP and the FPGA version of Handshake Join use internal memory on FPGA to store the tuples, which limits the maximum window size to several thousand, where PanJoin's subwindow on FPGA stores data in the external DDR3 memory which decides the maximum size of the window.

\section{PanJoin} \label{sec:pan}
\subsection{Architecture}  \label{sec:archtect}
PanJoin supports a sliding window theta join over two data streams (Stream $S$ and Stream $R$ in the following). PanJoin processes incoming data at two levels of parallelism: the \textit{architecture level}, or node level, where multiple worker nodes generate partial results in parallel, and \textit{thread level}, where each worker node executes with multiple threads.

\begin{figure}[!htb]
    \centering
    \begin{minipage}[b]{.40\textwidth}
  \centering
  \includegraphics[width=\textwidth]{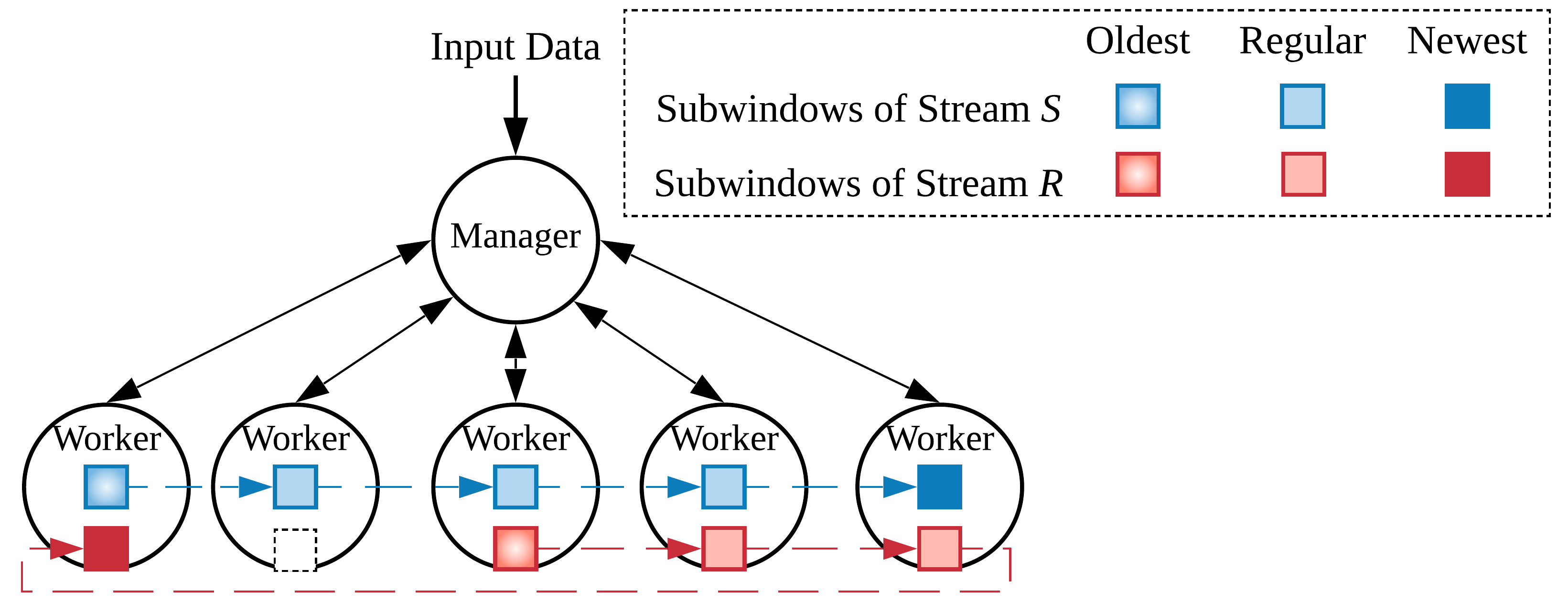}
  \caption{System architecture.}
  \label{fig:sys_arch}
    \end{minipage}\hfill%
    \begin{minipage}[b]{0.40\textwidth}
  \centering
  \includegraphics[width=\textwidth]{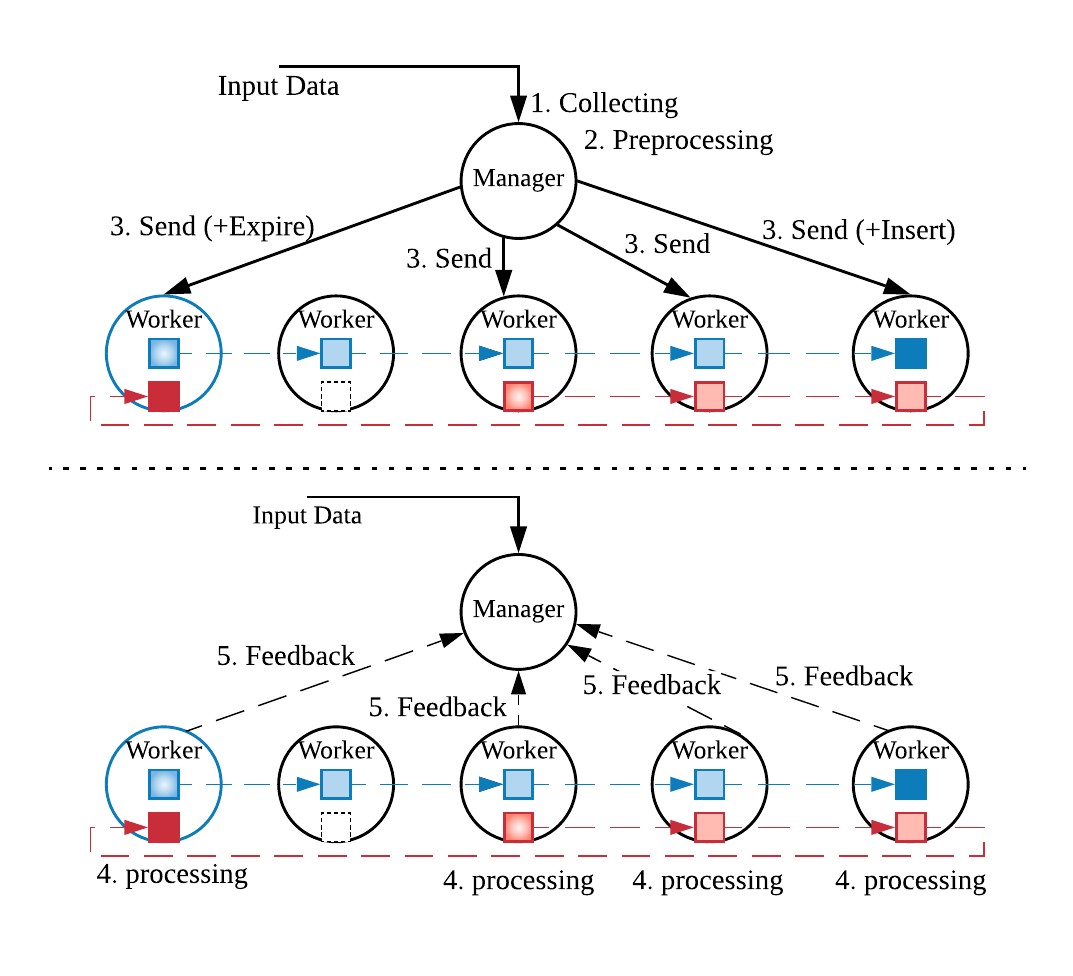}
  \caption{Process procedure. The buffer of Stream $S$ (blue) is full and is processed.}
  \label{fig:sys_process}
    \end{minipage} \hfill%
\end{figure}

The architecture of PanJoin is shown in Figure \ref{fig:sys_arch}. There are several worker nodes and one manager node. Each worker node has some number of subwindows. The subwindows of the same stream are chronologically chained and allocated in a round robin way among the worker nodes, and there is no communication between worker nodes, which is different from (Low Latency) Handshake Join \cite{teubner2011soccer, roy2014low}. The manager node manages the topology and the locations of all subwindows. The manager node also preprocesses incoming raw data and distributes the preprocessed data together with processing commands to the worker nodes. The processing commands includes: \textsl{create} subwindow, \textsl{insert} tuples, \textsl{probe} subwindow, and \textsl{expire} subwindow. 
% The processing commands instruct the corresponding worker nodes to perform the following operations:
% \begin{itemize}
% \item Create subwindow: create a new subwindow for a specified stream (Stream $S$ or $R$).
% \item Insert tuples: insert tuples into the specified subwindow and check if the subwindow becomes full.
% \item Probe: probe all the subwindows of the opposite stream of the incoming tuples and sent the result back to the manager node if necessary. 
% \item Expire: check if the entire subwindow should be expired %(and thus becoming empty). 
% \end{itemize}
To generate correct commands, the manager node needs to collect the running status from every worker node. This information is sent together with the result data from the worker nodes to the manager node with very low overhead (several bits). The two most important messages are whether the oldest subwindow is empty and whether the newest subwindow is full. Additionally, because the manager node needs to distribute data to worker nodes, to improve network and processor utilization, PanJoin can process more than one tuple packed as a batch simultaneously, which is referred to as \textit{batch mode}. 

%PanJoin can process one single tuple in \textit{single mode} or a batch of tuples in \textit{batch mode}. Because the manager node needs to distributed data to worker nodes, PanJoin uses batch mode to improve network and processor utilization. 

The high-level algorithm of PanJoin implements a five-step procedure, as demonstrated in Figure \ref{fig:sys_process}. Each of the steps is defined as follows:

\begin{enumerate}[label=Step \arabic*]

\item \textit{Collecting}: the manager node collects tuples and places them into two independent buffers, one for each stream.

\item \textit{Preprocessing}: The manager node retrieves the joining field from the incoming tuples and other fields that are necessary for the theta join. In batch mode, the manager node sorts the tuples by the values of the joining field. Then, the manager node decides whether to create a new subwindow if the current newest subwindow is full, as well as whether the oldest subwindows have to expire tuples. Subsequently, the manager node generates processing commands and packs the commands with the incoming tuples into messages.

\item \textit{Sending}: The manager node sends the messages to the worker nodes and ensures that the \textsl{probe} commands are sent to all the worker nodes with nonempty subwindow(s) of the opposite stream. In Figure \ref{fig:sys_process}, when the system is processing tuples from Stream $S$, the second worker node does not have a nonempty subwindow of stream $R$. Thus, it has no need to perform probing, and it will not receive a \textsl{probe} command. The message sent to the newest subwindow has an \textsl{insert} command, and the message sent to the oldest subwindow has an \textsl{expire} command. If necessary, the manager node sends a node with a \textsl{create} command to create a new subwindow. 

\item \textit{Processing}: The worker nodes receive messages and perform the processing commands. 

\item \textit{Feedback}: The worker nodes send the probing result (optional, depending on the topology of the processing nodes, e.g., worker nodes may directly forward their result to the nodes that process the ``select'' operation) and their running status back to the manager node. 

\end{enumerate}

Note that the manager node can perform Steps 1 and 2 in parallel with Steps 4 and 5 when the worker nodes are processing the data. We can also add one or several prefilter nodes ahead of the manager node to retrieve the joining field from the incoming tuples. In our preliminary experiment, the manager node can process more than 300M tuples per second, and the main bottleneck is in Step 4 in the worker nodes and network communication between the manager and worker nodes. To accelerate Step 4, we introduce three data structures to manage tuples inside a subwindow: \textit{Range Partition Table} (RaP-Table), \textit{Wide B$^+$-Tree} (WiB$^+$-Tree), and \textit{Buffered Indexed Sort} (BI-Sort). RaP-Table and WiB$^+$-Tree perform well when batch size and selectivity is small, while BI-Sort runs faster when batch size or selectivity is large. WiB$^+$-Tree is slower but more powerful than RaP-Table because it can handle increasing values. Users can choose one of these data structures for a subwindow, where the chosen data structure serves as an ``index'' and further divides the subwindow into partitions. Therefore, in the probing step, rather than scanning the whole window, the worker node only needs to check the tuples in a limited number of partitions, which significantly reduces the number of comparisons and improves the system throughput by several orders of magnitude. The following subsections provide detailed discussions about these data structures.

\subsection{RaP-Table}

RaP-Table range partitions the tuples in a subwindow. The partition boundaries, which are called \textit{splitters}, are stored in the \textit{partition table}. To find the target partition during insertion and probing, RaP-Table performs a binary search on the partition table. 
The main challenge for using range partitioning is handling the skew in real-time data, i.e., the data may unevenly distribute among the partitions according to the current splitters such that the subwindow needs more time to scan some large partitions to obtain the join result. In addition, data skew introduces another challenge for storing the tuples because RaP-Table also attempts to store the tuples continuously in the same partition to accelerate probing, and skewed data make it difficult to predict the proper size of each partition to allocate in memory. %, since any of the partitions may have to hold all the tuples in the subwindow if the data is highly skewed and concentrates to a tiny range.

RaP-Table provides a solution to the two challenges by implementing an \textit{adjustment algorithm} and a data storage structure called \textit{Linked List Adaptive Table} (LLAT). 

\subsubsection{Adjustment Algorithm} \label{adjust_algo}

When a new subwindow is created, it receives a new partition table calculated based on the sampling information in its predecessor. The sampling information includes three histograms: the tuple count $count_i$, the maximum tuple value $max_i$, and the minimum tuple value $min_i$ of each partition. %The tuple count decides the offset of the adjustment in its corresponding partition boundary, while the maximum value and the minimum value determine the actual value of the partition boundary. 
The main idea is to scan the histogram of the tuple count to find an approximate range for each splitter and use two other histograms to calculate a new value for each splitter.

\begin{figure*}[!htb]
    \centering
    \begin{minipage}[b]{.26\textwidth}
  \centering
  \includegraphics[width=\textwidth]{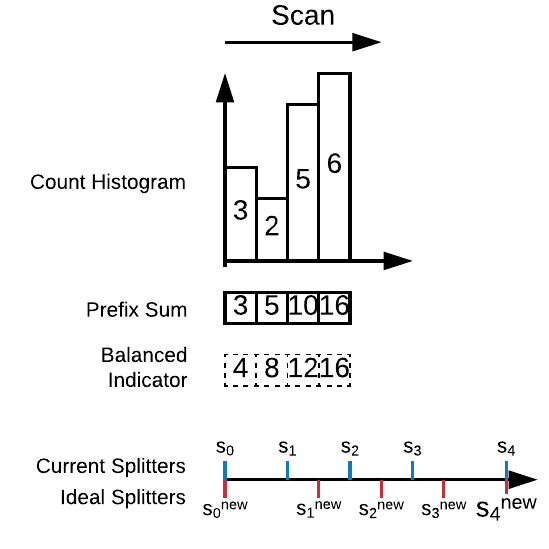}
  \caption{Scan the count histogram.}
  \label{fig:adjust1}
    \end{minipage}\hfill%
    \begin{minipage}[b]{0.26\textwidth}
  \centering
  \includegraphics[width=\textwidth]{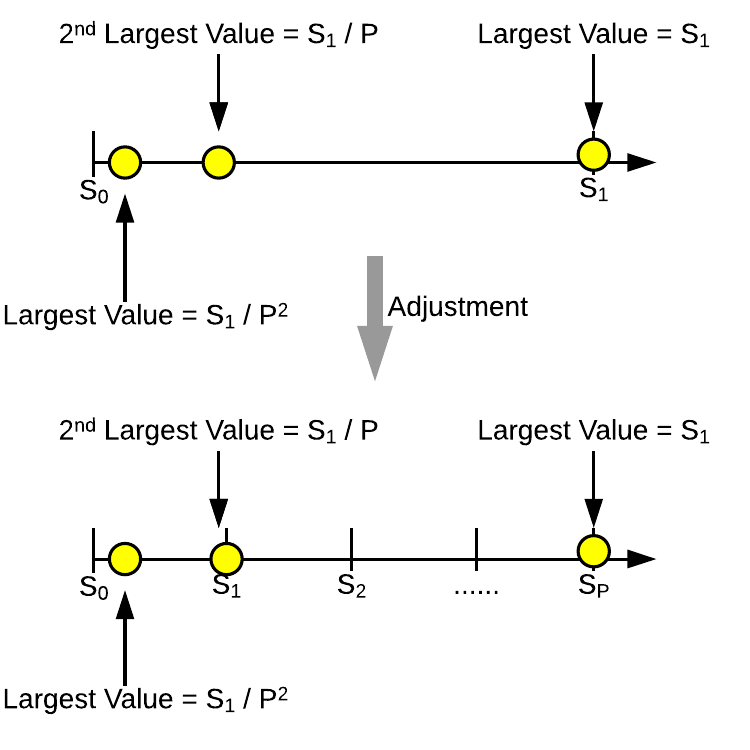}
  \caption{A adjustment worst case.}
  \label{fig:adjust2}
    \end{minipage} \hfill%
    \begin{minipage}[b]{0.26\textwidth}
  \centering
  \includegraphics[width=\textwidth]{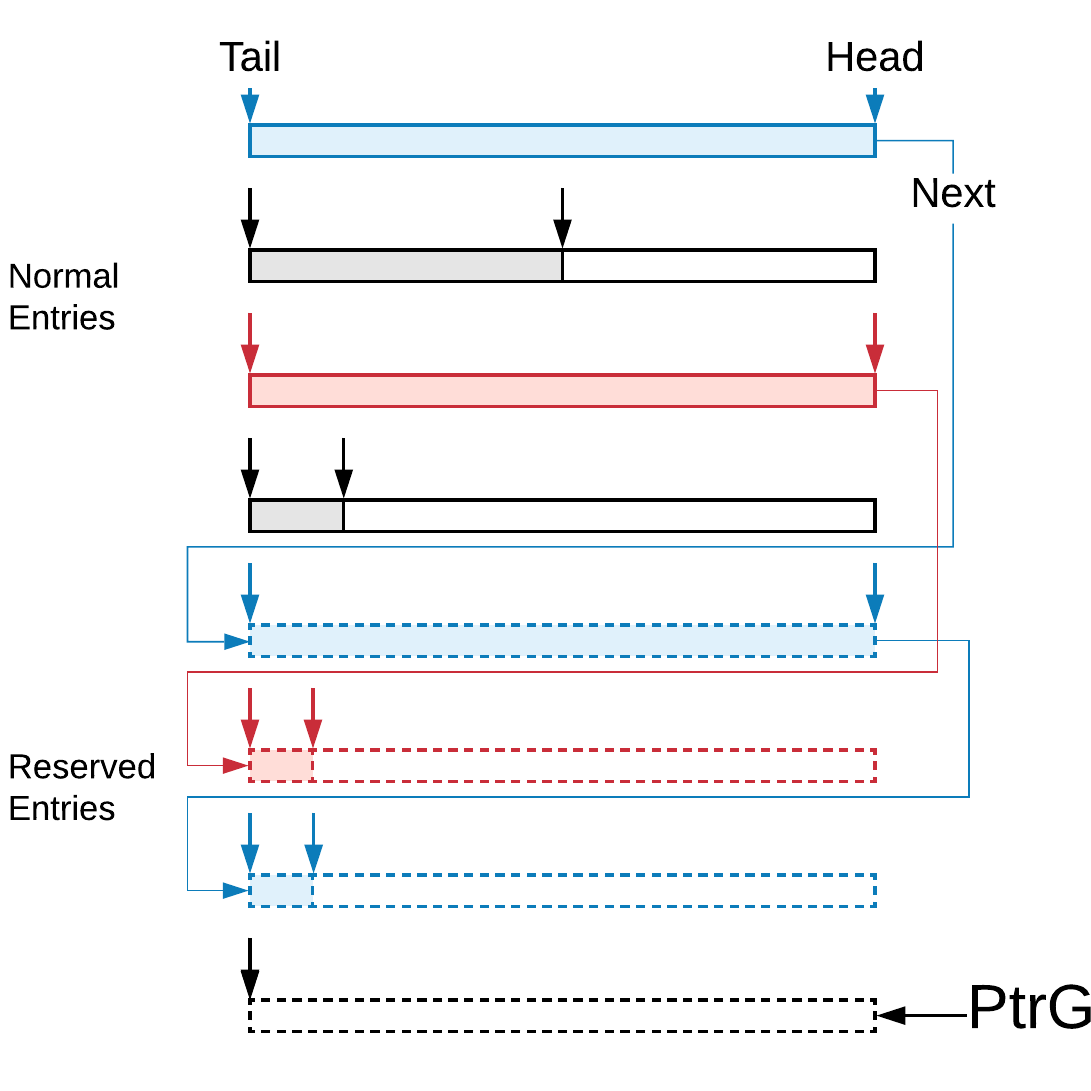}
  \caption{An LLAT with $P = 4$.}
  \label{fig:llat}
    \end{minipage}
\end{figure*}

First, we calculate the prefix sums $sum_i = sum_{i-1} + count_i$ of the tuple count histogram and the \textit{balancing indicator} $bal_i = N / P \cdot i $, where $N$ is the total tuple count and $P$ is the partition count. The value of $bal_i$ is the ideal value of $sum_i$ if the tuples are evenly partitioned. In the example shown in Figure \ref{fig:adjust1} where $N = 16$ and $P = 4$, we have $bal_2 = 8$, which means that the previous 2 partitions should have 8 tuples in the ideal case. For any two partitions $i$ and $j$, if $bal_j \in (sum_{i-1}, sum_i]$, we know that the new $j^{th}$ splitter $s^{new}_j$ should be the value of a tuple stored in the $i^{th}$ partition, i.e., $s^{new}_j \in [min_i, max_i]$. For example, in Figure \ref{fig:adjust1}, $bal_2 = 8 \in (sum_2, sum_3] = (5, 10]$, which indicates that the $s^{new}_2$ should be a tuple value stored in the $3^{rd}$ partition, i.e., $s^{new}_2 \in [min_3, max_3]$.

Then, we compute the value of $s^{new}_j$. Since we do not have more information about the distribution inside a partition, we assume that they follow a uniform distribution. Thus, according to the feature of a uniform distribution, the value of $s^{new}_j$ should be:
$$
s^{new}_j = \frac{bal_j - sum_{i-1}}{count_i} \cdot (max_i - min_i) 
$$
The calculation of the new splitters can be performed in a single loop. We present the pseudocode in Algorithm \ref{alg:adjust}. For each iteration, we compute the prefix sum $sum_i$ (Line 4). Because there may be more than one $bal_j$ that satisfies $bal_j \in (sum_{i-1}, sum_i]$, once we find a valid $bal_j$ (Line 5), we use a while loop to check all possible $bal_j$ values (Line 6) and  calculate the new splitters (Line 7). The time complexity of the algorithm is $\mathcal{O}(P+P) = \mathcal{O}(P)$ because we basically generate and scan the two arrays ($sum_i$ and $bal_i$) of size P only one time.

%Note that we do not need a nested loop to scan $j$s: if $accu_K >= in_i$, all $accu_j$ that $j>K$ is larger or equal to $in_i$. Thus, for each iteration, we started from the $accu_K$ that was left by the previous iteration so that $accu_K >= in_{i-1}$. We then check $accu_{K+1}$, $accu_{K+2}$ ... until $accu_{j} >= in_{i}$, and leave $accu_{j}$ for the next iteration. 

\begin{algorithm}[ht]
\KwData{Histogram $count_i$, $max_i$, $min_i$, $i$ in $1...P$}
\KwResult{New splitters $s^{new}$}
\Begin{
$sum_0 \longleftarrow 0$; $bal_1 \longleftarrow N / P$; $i \longleftarrow 0$; $j \longleftarrow 0$\;
\While{$sum_i < N$}{
	$i \longleftarrow i + 1$; $sum_{i} = sum_{i-1} + count_i$\;
    \If{$bal_j \in (sum_{i-1}, sum_i]$}{
      \While{$bal_j <= sum_i$}{
        $s^{new}_j \longleftarrow \frac{bal_j - sum_{i-1}}{count_i} \cdot (max_i - min_i)$\;
        $j \longleftarrow j + 1$; $bal_{j} \longleftarrow bal_{j-1} + N / P $\;
      }
    }
}
}
\caption{Splitter Adjustment}
\label{alg:adjust}
\end{algorithm}

For this algorithm, we have constructed a worst case scenario, as shown in Figure \ref{fig:adjust2}. In this case, before calling the adjustment algorithm, all of the data are inserted into one partition (for convenience, let us assume that it is the $1^{st}$ partition). The data are distributed inside that partition as follows: the largest value equals the splitter $s_1$, the second largest value equals $s_1/P$, the third largest value equals $s_1/P^2$, and so forth. Now, because $min_1 = 0$, $max_1 = s_1$, $bal_i = N/P \cdot i$, $count_1 = N$ and $sum_0 = 0$, the algorithm provides $s^{new}_1 = s_1/P$, and other splitters $s^{new}_i = s_1 \cdot i/P$. However, in this way, all updated partitions are still empty, except the $P^{th}$ partition holds one tuple and the $1^{st}$ partition holds the remaining tuples. The same situation will repeat after another adjustment and will end once the the range $[min_1, max_1]$ is not divisible by $P$. If we assume that the values are 32-bit integers, then the maximum number of adjustments is $\lceil log_{P}2^{32}\rceil$. If $P = 1024$, then $\lceil log_{P}2^{32}\rceil = 4$. If $P = 65536$, then $\lceil log_{P}2^{32}\rceil = 3$. Thus, the algorithm needs only a few calls to adjust the splitters. In Section \ref{sec:eva_rap}, we show that for some commonly used distributions, the algorithm can converge in only 1-3 times of adjustment.

The size of a histogram equals the partition count $P$. RaP-Table uses three histograms (count, maximum value, and minimum value) and a partition table, which are four arrays that hold $4P$ elements in total. An element has the same size as a tuple value. Since the partition count is considerably smaller than the subwindow size $N_{Sub}$, e.g., $P=64K$ and $N_{Sub}=8M$, the overhead of memory allocation for those arrays is negligible ($64K * 4 / 8M < 0.04$).

\subsubsection{LLAT}

RaP-Table requires a storage strategy that keeps the memory access pattern continuous and adapts to the data skew.
To meet these requirements, we designed a data structure called \textit{Linked List Adaptive Table} (LLAT). Figure \ref{fig:llat} illustrates the structure of an LLAT. Each LLAT has $2P$ entries, where $P$ is the partition count. The first $P$ entries, the \textit{normal entries}, are used for all partitions, with one entry per partition. The last $P$ entries are called \textit{reserved entries} which store skewed data. The LLAT has a global pointer \textsl{PtrG} that points to the first unused reserved entry. Each entry has an array that holds $(N_{Sub} / P) \cdot \sigma$ tuples belonging to the same partition, where $N_{Sub}$ is the subwindow size and $\sigma$ is a threshold larger than 1. We suggest that $\sigma$ should be approximately 1.10-1.25 to handle the small skew in the real-time data. Each entry has two data pointers: \textsl{Head} and \textsl{Tail}. Similar to the case in a linked list, each entry also has a \textsl{Next} pointer referring to an entry in the LLAT.

The insertion and deletion operation is defined as follows. Initially, in every entry, the \textsl{Head} and the \textsl{Tail} point to the zero position of its array. The \textsl{PtrG} is set to $P$. Once a tuple obtains its partition index $i$, it is inserted at the partition's \textsl{Head}, which is then increased by 1. If an entry is full after an insertion, its \textsl{Next} pointer will point to the entry that \textsl{PtrG} refers to, then \textsl{PtrG} is increased by 1. In this way, we create a linked list for an unbalanced partition. Once an unbalanced partition receives another new tuple, LLAT will go through its linked list and insert the tuple at the last entry. The deletion is straightforward. When the LLAT receives the expired request with a value of a partition index, it increases the \textsl{Tail} of the corresponding entry by 1. If the entry is empty (its \textsl{Head} equals its \textsl{Tail}), LLAT will check the entry indicated by the \textsl{Next} pointer. This process continues iteratively until LLAT finds a nonempty entry and expires the tuple. 

It is easy to prove that $2P$ entries are enough for all the cases. Suppose that we have a case in which $2P$ entries are not enough. When this occurs, there must be at least $P$ entries that are full (so they require $P$ reserved entries). Because $\sigma$ is larger than 1, it means that the subwindow has at least $(N_{Sub} / P) \cdot \sigma \cdot P = N_{Sub} \cdot \sigma > N_{Sub}$ tuples, which is not possible because a subwindow has at most $N_{Sub}$ tuples.

The storage overhead of LLAT can be calculated as
$(N_{Sub} / P \cdot \sigma) \cdot 2P / N_{Sub}-1$. When $\sigma$ is 1.25, the overhead equals 150\%, meaning that we need additionally 1.5 times more space to store all the data, or the space utilization is 40\%, which is acceptable in practice.

\subsubsection{Known Issues}

RaP-Table has a limitation in handling incoming data that have increasing values. For example, the id field of the tuples generally contains increasing values. Because RaP-Table generates the new partition table based on the data pattern in the previous subwindow, the new partition table is not able to partition newly arrived tuples evenly but delivers them into one or few partitions. For example, if the value range of the previous subwindow is $[0, 1000]$, the generated new partition table will attempt to partition the values between $[0, 1000]$. If the following data has a value range of $[1000, 2000]$, the new partition table becomes powerless. 

To address this issue, we designed a new method based on B$^+$-tree called \textit{Wide B$^+$-Tree} (WiB$^+$-Tree), a slower but more powerful data structure, which is shown in the following subsection.

\subsection{WiB$^+$-Tree}

\textit{Wide B$^+$-Tree} (WiB$^+$-Tree) is a data structure adapted from regular B$^+$-tree; therefore, it naturally adapts to skewed data. A leaf node in WiB$^+$-Tree is designed similar to a partition in RaP-Table, while internal nodes are used as the partition table to index the leaf nodes. The major differences between a WiB$^+$-Tree and a B$^+$-tree is that the leaf nodes have different configurations than the internal nodes in a WiB$^+$-Tree. 

First, a leaf node has more elements than internal nodes. We would like to keep the size of the internal nodes small enough such that the internal nodes remain in the CPU cache. %, while keeping the size of the leaf nodes large enough to utilize memory bandwidth. 
An internal node with 64 elements of 32-bit integers needs at least 256 Bytes of memory, which means that a modern CPU core is able to hold 4096 nodes in its private 1 MB L2 cache or approximately 85,000 nodes in the shared 22 MB L3 cache, which are enough for a WiB$^+$-Tree to index a large subwindow (larger than 1M).
%The size of the leaf nodes depends on the memory access latency and memory bandwidth.
%because our subwindow can be large (millions of tuples), there is not enough space to hold all tuples of in the CPU cache, thus we have to store the leaf nodes in the main memory. 
%In practice, each read or write to a leaf node requires a random memory access. Suppose the access penalty is 150 cycles. During this time, we can linearly scan 150 tuples in a fully pipelined fashion on CPU. A DDR-2400 memory can deliver a 64-bit tuple at 2400MHz, which is approximately the same frequency a modern CPU has. In this case, a leaf node should have more than 450 tuples so that we can obtain a memory bandwidth utilization higher than $450 / (150+450) = 75\%$. Therefore, we want to make leaf nodes large enough to amortize the access latency.
%Therefore, we have to perform a large amount of random access to main memory during the comparison phase of join processing. %and may scan more if the implementation is empowered by SIMD

Second, the elements in the leaf nodes are unsorted, unlike the internal nodes where all the elements are sorted. We sort the leaf node only when we need to split it into two nodes. 
Suppose that the width of a full leaf node is $W$; if we keep the order inside the leaf node, the time complexity of obtaining a full node (inserting $W$ elements in that node) is $\mathcal{O}(W^2)$, while a sort operation at the time of node splitting has a complexity of only $\mathcal{O}(WlogW)$.
In our preliminary experiment, we found that a WiB$^+$-Tree with unsorted leaf nodes commonly has 3-5 times less insertion time than a WiB$^+$-Tree with sorted leaf nodes.    

Third, no internal node has duplicate elements. All the tuples with the same value will be inserted into the same leaf node, which may cause some leaf nodes to have more elements than their maximum width. To efficiently store the extra tuples, we use LLAT to organize all the tuples stored in all the leaf nodes. A slight difference between the LLAT in WiB$^+$-Tree and RaP-Table is that the \textsl{PtrG} pointer initially points to the first partition, since there is no leaf node serving as a partition when the subwindow is empty. Each leaf node holds a pointer to an entry in the LLAT,  while the other information of the leaf node, such as node size, can be held in the CPU cache.
%Note that in the real implementation, we put all the nodes into arrays and implement pointers to a node as a array index. In this way, we avoid to use expensive dynamic memory allocation calls. 

\subsection{BI-Sort}\label{sec:bisort}

After we designed RaP-Table and WiB$^+$-Tree, we found that both data structures sort the subwindow at a coarse level. RaP-Table behaves similar to a bucket sort, while WiB$^+$-Tree indexes those ``buckets" with a heterogeneous B$^+$-Tree. Thus, we design a data structure that genuinely sorts the entire subwindow at a fine level.

There are two challenges in designing a sorted-based data structure. First, we have to spend $\mathcal{O}(N)$ time to insert a new tuple into the correct position. Although we can find the address in $\mathcal{O}(logN)$ time through a binary search, we still need $\mathcal{O}(N)$ time to shift the tuples larger than the inserted tuple to the new addresses. 
Second, in the probing operation, even though we can find the target tuple in $\mathcal{O}(logN)$ time with a binary search, each step of the binary search has a memory access that can be treated as a random memory access. Although the first few accesses may hit the cache, when $N$ becomes large, e.g., 8M = $2^{23}$, most of the memory accesses have a considerable delay, which slows the probing operation. %Suppose the memory access latency is 40 ns, a probe access need requires $40 \cdot 23 = 920$ns, which equals a maximum throughput of only 1.08M/s, which is \hl{30 times} slower than BI-Sort can offer.  

BI-Sort overcomes the first challenge with an \textit{insertion buffer}. When a new tuple arrives, it is first inserted into the insertion buffer. The insertion buffer has a limited size $B$, and all the data are unsorted. The new tuple will remain in the buffer until the buffer is full, then the buffer will be sorted and merged into the subwindow data that are stored as a sorted array called the \textit{main array} in the memory. Merging two sorted arrays requires $\mathcal{O}(M+B)$ time ($M$ is the size of the main array), which is shared by $B$ tuples. Therefore, we can significantly reduce the insertion time such that the insertion will not be the bottleneck of the stream join processing. When the subwindow is probed, both the tuples in the main array and the insertion buffer will be both probed. Since the insertion buffer has a limited size, it can be stored in the cache to have a limited time cost of a linear scan.

To address the second challenge, BI-Sort adds an \textit{index array} for the main array. The index array is updated immediately after the insertion buffer is merged into the main array. The index array samples the value of every $M/P$ tuple in the main array, where $P$ is the size of the index array, i.e., the partition count. The memory space between two adjacent indexed tuples is called a \textit{partition}. When $P$=64K, the index of 32-bit integer needs only 256 KB, which can easily be stored in the L2 private cache. When BI-Sort needs to perform a binary search on the main array, it first performs the binary search on the index to find the target partition, and then it performs a binary or linear search (batch mode, discussed in the following subsection) inside the partition. Note that usually $P << M$; therefore, the time complexity of merging the main array together with updating the index is $\mathcal{O}(M+B+P) = \mathcal{O}(M+B)$. Thus, updating index has little effect on the performance.
%In our preliminary experiments, we found that linear search is fast enough for $P < 255$. This is because current DDR4 memory usually has a minimum memory access latency (CAS latency) about 10ns and a transfer time (a 64 bit word) about 0.4ns. Thus, during the latency of a random read of a word, the DDR memory can transfer 25 continuous words. 

In conclusion, to overcome the disadvantages of the traditional sort-based solution, we add a buffer and an index to a ``merge sort''-based data structure, which is named \textit{Buffered Indexed Sort}. 
%BI-Sort offers several advantages over RaP-Table and WiB$^+$-Tree. First, 
BI-Sort has the smallest storage overhead among the three data structures: the size of the main array is the same size as the subwindow, while RaP-Table and WiB$^+$-Tree use LLAT that needs more space than BI-Sort. With the same number of partitions, the index array and the insertion buffer of BI-Sort also require less space than the partition table and histograms in RaP-Table, as well as the tree nodes in WiB$^+$-Tree. 
%Second, the algorithm of BI-Sort is the simplest and has the highest throughput. 
BI-Sort is also the simplest algorithm, which allows us to implement it on FPGA. 
Furthermore, because BI-Sort saves the tuples in order, the theta join result of a probing tuple can be represented as a $<\mathsf{id_{start}}, \mathsf{id_{end}}>$ record (with a label \textit{not} for condition $\neq$), where $\mathsf{id_{start}}$ is the index of the main array where the result starts and $\mathsf{id_{end}}$ is the index where the result ends. If the join condition has more than one band, e.g., $a \in [b-5, b+5] \lor a \in [b+20, b+35]$, BI-Sort uses the same number of records as the bands to represent the result. Therefore, when the selectivity is very high (e.g., most of the tuple values in the stream are the same), RaP-Table and WiB$^+$-Tree need to copy a large amount of tuples into the result, while BI-Sort simply returns index records. In this way, BI-Sort can significantly reduce the size of the feedback message to the manager node. To support this, the manager node has to maintain a mirror of the main array of every subwindow. The main array in the manager node will perform the same insertion as in the corresponding worker node. In our preliminary experiment, there is no performance overhead to keep a mirror of each subwindow, since the insertion in the manager node is performed in parallel with the newest worker node.

\subsection{Batch Mode}

\begin{figure}
  \centering
  \includegraphics[width=.40\textwidth]{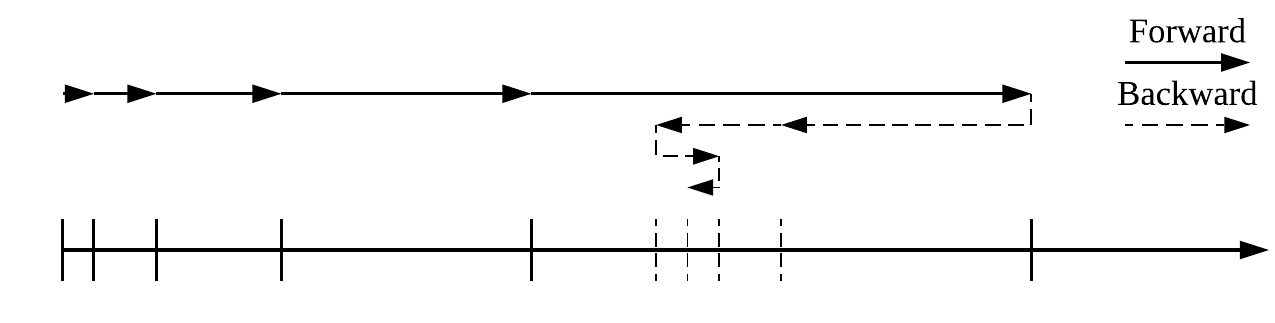}
  \caption{Rebounding binary search.}
  \label{fig:rebounding}
\end{figure}

In PanJoin, the manager node needs to send incoming data to the worker nodes. To fully utilize the network bandwidth, PanJoin supports \textit{batch mode} where a batch of tuples will be processed simultaneously. In addition to batching the tuples at the architecture level, all three of the data structures support batch mode for insertion and probing.

RaP-Table uses a modified partitioning algorithm for  insertion and probing in batch mode. Rather than regular binary search, RaP-Table uses a \textit{rebounding binary search}, shown in Figure \ref{fig:rebounding}. Rebounding binary search has two phases: forward phase and backward phase. Starting with an initial element in the sorted array, the algorithm goes in one direction in the forward phase, then  turns back to backward phase if it reaches a value larger than the target value, and then performs a regular binary search. The step length is doubled after each comparison in the forward phase and halved in the backward phase. Because all the tuples in the batch are presorted by the manager node, the target partition for the next tuple must have an index larger or equal to the target partition of the current tuple. Therefore, during insertion and probing, RaP-Table uses rebounding binary search to find the target partitions of the batched incoming tuples : each tuple starts it search from the partition id of its previous tuple, which is faster than performing a ``complete'' binary search per tuple. In this manner, the partition phase only accounts for less than 10\% of the total execution time of the insertion and probing operation, which is competitive to a hash-based solution, e.g., SHJ. When probing the subwindow, the tuples with the same target partition can share the same memory access, which also notably improves the performance. 

When traversing the tree, WiB$^+$-Tree starts with the node path in the tree of the previous batch tuple, backtracks to the upper levels, and then goes downwards to the correct leaf node if necessary. During probing, similar to RaP-Table, WiB$^+$-Tree first finds the target partitions for all the batch tuples, and then it scans the partitions to improve the performance. Both RaP-Table and WiB$^+$-Tree use a nested loop (inner: related batch tuples, outer: partition tuples) to scan a target partition of several related batch tuples. 

BI-Sort first compares the sizes of the batch and the insertion buffer before actual insertion. If the batch is larger than the buffer, BI-Sort directly merges the batch into the main array. Otherwise, it places all the batch tuples into the insertion buffer. While performing probing, BI-Sort first uses the rebounding binary search to find the target partitions of the batch tuples, and then it scans the corresponding partitions. Because the tuples in the partition and related batch tuples are both sorted, the result index of tuple $t_i$ must be larger than or equal to the result index of $t_{i+1}$. Thus, rather than a nested loop, BI-Sort can simply scan the partition with a merged-like loop: if the current partition tuple is smaller than the join condition of the current batch tuple, we increase the iterator of the partition tuples; otherwise, we increase the iterator of the batch tuples. In this way, each partition tuple is accessed only once: thus, BI-Sort can achieve higher performance than the other two data structures.  
We illustrate this algorithm by using equi-join as an example. In RaP-Table and WiB$^+$-Tree, because the partition is unsorted, a partition tuple in partition $P$ needs to be joined with all the batch tuple with the same target partition as $P$. In BI-Sort, when a batch tuple $t_1$ reaches to a partition tuple with a larger value than $t_1$, it stops probing and this partition tuple becomes the start of the next batch tuple $t_2$. Here, we assume $t_1$ has larger value than $t_2$. Otherwise, $t_2$ will have the same value as $t_1$ (batch is sorted), then BI-Sort copies the join result from $t_1$ to $t_2$. In this way, a partition tuple in partition $P$ only needs to be joined with a single batch tuple. Therefore,  Although RaP-Table and WiB$^+$-Tree can sort their partitions after the subwindow is full and stays unchanged, this can cause some extra overhead and may significantly reduce the performance in some cases. Thus, in this paper, all the partitions in RaP-Table and WiB$^+$-Tree remain unsorted. 

In practice, the manager node sets two conditions of maximum collecting time and maximum tuple count for batch mode: either of the two conditions is satisfied, the manager node packs all the collected tuples into a batch and starts processing. 

%In BI-Sort, a batch tuple only joined with the partition tuples with higher index than those joined by the previous batch tuple.

\subsection{Equi-Join vs. Non-Equi-Join} \label{sec:join_con}

The three data structures each have different strategies for equi-join and non-equi-join to probe partitions.

\subsubsection{Equi-join} In RaP-Table and WiB$^+$-Tree, each probing tuple has a single target partition. In BI-Sort, a probing tuple may have multiple target partitions because tuples with the same value may be stored across several partitions. Therefore, in BI-Sort, the rebounding binary search provides the target partition $P_i$ with the lowest id $i$. Then, BI-Sort converts the current equi-join $x=v$ into a non-equi-join as $x \in [v, v^+)$, where $v^+$ means the smallest value larger than $v$.

%starts probing with $P_i$. If the last tuple in $P_i$ has the same value of the probing tuple, we know that $P_{i+1}$ may has some tuples with the same value. Next, BI-Sort probes $P_{i+1}$ and the same procedure repeats until the last tuple of $P_{i+k}$ has a different value. Note that in batch mode cross-partition probing does not generate any overhead because every tuple in the subwindow is still joined with a single probing tuple.

\subsubsection{Non-equi-join} Except for the condition $\neq$, which can be performed by a equi-join with a filtering operation, PanJoin will calculate the upper bound and lower bound of the band of the join condition for each probing tuple. Then, three data structure probes the target partitions $P_{low}$ of the lower bounds, and the target partition $P_{up}$ of the upper bounds if $P_{low} \neq P_{up}$. Additionally, RaP-Table and WiB$^+$-Tree copy the partitions between $P_{low}$ and $P_{up}$, while BI-Sort can skip this procedure if it chooses $<\mathsf{id_{start}}, \mathsf{id_{end}}>$ as its result format.
%RaP-Table first probes the target partition of the lower bound $P_{low}$. If the upper bound has a different target partition $P_{up}$, RaP-Table also probes $P_{up}$, and then directly copies the tuples from the partitions between $P_{low}$ and $P_{up}$. WiB$^+$-Tree has a similar procedure, except that it has to traverse the tree to find the partitions that hold values between $P_{low}$ and $P_{up}$. On the other hand, if BI-Sort chooses $<\mathsf{id_{start}}, \mathsf{id_{end}}>$ as the result format, it only probes the $P_{low}$ and maybe $P_{up}$ to find the index $\mathsf{id_{start}}$ and $\mathsf{id_{end}}$ without copying the tuples in the middle partitions. Otherwise, BI-Sort uses the same procedure as RaP-Table.

\subsection{Miscellaneous Implementation Decisions} 

%There are several implementation decisions in PanJoin that we will discuss in detail. 

\subsubsection{Expiration} PanJoin expires an entire subwindow (the oldest one) instead of several tuples in the oldest subwindow. Therefore, none of the three data structures currently has a deletion operation (LLAT has, but RaP-Table does not). Since there will be a number of subwindows for a stream and all of them are probed in parallel, an extra subwindow will not cause much overhead. While probing the oldest subwindow, we employ a filtering operation to remove expired tuples in the result. 

\subsubsection{Count-based Window vs. Time-based Window vs. Out-of-Order Window} A time-based window needs extra fields in each tuple to save the event time or arrived time, depending on the application requirements. There is no difference in using the three data structures. The running status of each worker node saved in the manager node is slightly different: for a count-based window, the manager node monitors the tuple count in each subwindow, while for a time-based window, the manager node also saves the time fields of the oldest and newest tuple in each subwindow. These fields are used for expiring the subwindow: when all the tuples are older than the watermark, the entire subwindow is expired. This mechanism can also handle out-of-order tuples. In addition, when there are a few late arrived tuples in a subwindow, the subwindow can put them into a small buffer instead of its data structure so that when there is a probing request, it does not need to probe the whole data structure to find the join result. 

\subsubsection{Multithreading in Worker Nodes} When we implement the three data structure in batch mode, we find that a single thread on CPU cannot fully utilize the memory bandwidth. Therefore, in batch mode, all of the three data structures perform probing with multithreading. The workload is divided based on how many partitions each thread need to probe. In addition, BI-Sort also performs insertion with multiple threads, where BI-Sort tries to balance the size of each piece of the partially merged main array. %We do not parallelize insertion in RaP-Table and WiB$^+$-Tree because it does not affect the system performance of PanJoin significantly.  

\section{BI-Sort on FPGA} \label{sec:imp}

\begin{figure}[!htb]
    \centering
    \begin{minipage}[b]{.40\textwidth}
  \centering
  \includegraphics[width=\textwidth]{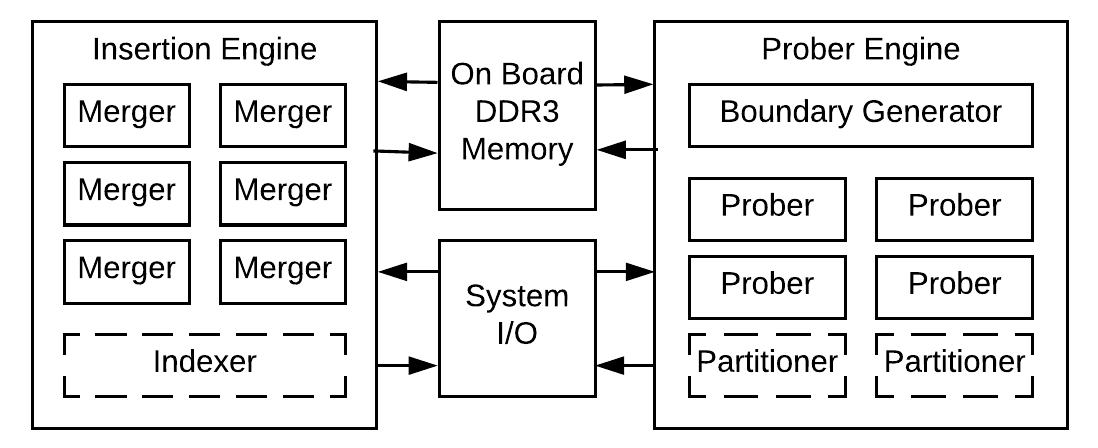}
  \caption{System architecture of FPGA solution.}
  \label{fig:fpga_arch}
    \end{minipage}
    \centering
    \begin{minipage}[b]{.20\textwidth}
  \centering
  \includegraphics[width=\textwidth]{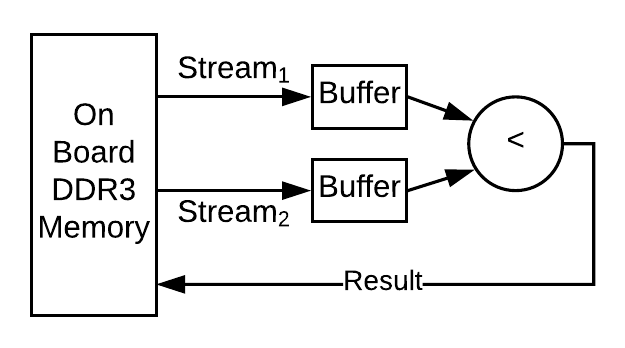}
  \caption{Merger.}
  \label{fig:merger}
    \end{minipage}
    \begin{minipage}[b]{0.20\textwidth}
  \centering
  \includegraphics[width=\textwidth]{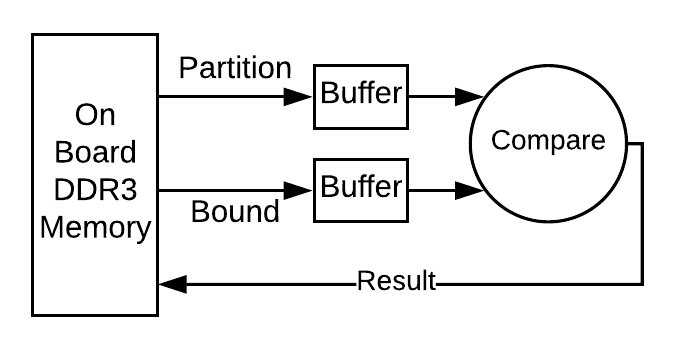}
  \caption{Prober.}
  \label{fig:prober}
    \end{minipage} \hfill%
\end{figure}

Because BI-Sort is simple enough to implement a hardware version, we attempt to build a worker node with BI-Sort on FPGA by Intel OpenCL in order to benefit from the two major advantages of FPGA: high throughput and low energy consumption. The system architecture is shown in Figure \ref{fig:fpga_arch}. The system has an insertion engine and a probing engine. Both of these engines can access the data buffered in the external memory on the FPGA board. The data can either come from the Internet I/O port that connects to the manager node or come from the host computer, depending on the configuration of the system.

The insertion engine has several \textsf{merger}s that merge the insertion buffer into the main array. 
%Each \texttt{merger} behaves like a thread in the CPU version of BI-Sort. 
The structure of a merger is shown in Figure \ref{fig:merger}, where there are two input stream ports, one output stream port, and a comparator. Here, we implement continuous memory access as a data stream to utilize the memory bandwidth. Initially, the merger reads one tuple from each input stream. Then, in each iteration, it compares the two tuples, chooses the smaller one to write into the output stream, and reads one tuple from the chosen stream. When all the mergers finish their work, an \textsf{indexer} will generate the index array. 

During probing, the \textsf{boundary generator} inside the probing engine first calculates the upper bound and lower bound of the join condition per probing tuple. Then, the \textsf{partitioner}s find the target partition(s) of each probing tuple. Subsequently, the \textsf{prober}s probe partitions and generate the result. The design of a prober is shown in Figure \ref{fig:prober}, which is similar to a merger. It has an input port for target partitions, an input port for bounds (upper or lower bounds), an output port for results, and a comparator. Initially, it reads a partition tuple and a bound. Then, in each iteration, if the value of the partition tuple matches the join condition, it writes $i$ into the result and reads the next bound. If the tuple exceeds the join condition, it reads the next bound. Otherwise, it reads the next tuple. % For example, for the upper bounds of a band join, if the tuple value is larger than the current upper bound, we say that the tuple value exceeds join condition, and vise visa. However, if the tuple is the last one that is smaller or equal to the current upper bound, it then match the condition of band join and the prober writes the id of the tuple into the result. 

\section{Evaluation} \label{sec:eva}

In this section, we first show the analytical performance of the three data structures on CPU and of BI-Sort on FPGA. Then, we show the throughput of PanJoin as a whole system and the comparisons with other stream join solutions. All the throughput axes in the following figures are in \textbf{log} scale.

\subsection{Analytical Evaluation} \label{sec:ana_eva}

\begin{figure*}[!htb]
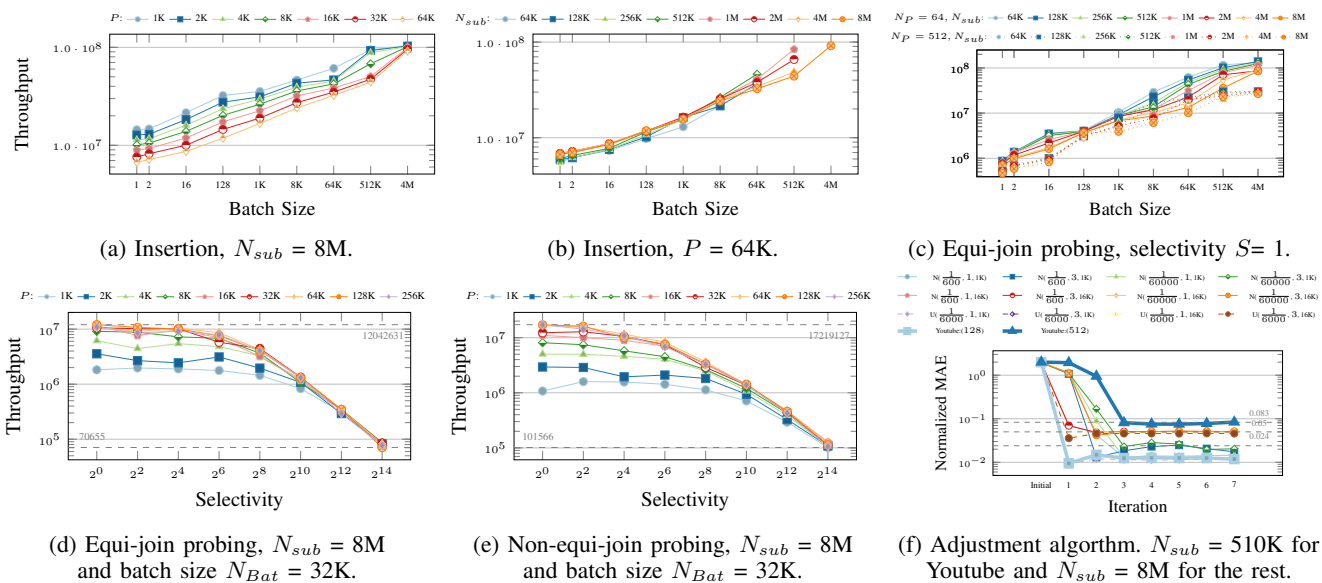

\captionsetup[subfigure]{justification=centering}
    \centering
    \begin{subfigure}[b]{0.32\textwidth}

\centering
\includegraphics[width=\textwidth]{figures/rap_insert_1.tikz}
\caption{Insertion, $N_{sub}$ = 8M.}
\label{fig:rap_insert_1}

    \end{subfigure}%
    \begin{subfigure}[b]{0.32\textwidth}

\centering
\includegraphics[width=\textwidth]{figures/rap_insert_2.tikz}
\caption{Insertion, $P$ = 64K.}
\label{fig:rap_insert_2}
    
    \end{subfigure}%
\begin{subfigure}[b]{0.32\textwidth}
\centering
\includegraphics[width=\textwidth]{figures/rap_probe_1.tikz}
\caption{Equi-join probing, selectivity $S$= 1.}
\label{fig:rap_probe_1}
    
    \end{subfigure}
        \begin{subfigure}[b]{0.32\textwidth}
\centering
\includegraphics[width=\textwidth]{figures/rap_probe_2.tikz}
\caption{Equi-join probing, $N_{sub}$ = 8M \\ and batch size $N_{Bat}$ = 32K.}
\label{fig:rap_probe_2}

\end{subfigure}
\centering
 \begin{subfigure}[b]{0.32\textwidth}
 \includegraphics[width=\textwidth]{figures/rap_probe_3.tikz}
\caption{Non-equi-join probing, $N_{sub}$ = 8M  \\ and batch size $N_{Bat}$ = 32K.}
\label{fig:rap_probe_3}
 \end{subfigure}
  \begin{subfigure}[b]{0.32\textwidth}
  \centering
   \includegraphics[width=\textwidth]{figures/rap_adjust.tikz}
\caption{Adjustment algorthm. $N_{sub}$ = 510K for Youtube and $N_{sub}$ = 8M for the rest.}
\label{fig:rap_adjust}
   \end{subfigure}
   \caption{Performance of RaP-Table.}
      \label{fig:rap}
\end{figure*}

We implemented the three data structures in C++17. The program is tested on one node of a high-performance cluster released in 2018. The node has two 20-core Intel\textsuperscript{\textregistered} Xeon\textsuperscript{\textregistered} Gold 6148 processors and 192 GB DDR4 memory with a total bandwidth of 150 GB/s. We evaluate the performance of the insertion and probing operation for each data structure by measuring the throughput of input tuples (unit: tuples per second) on the host subwindow of the data structure. Each input tuple has a format of $<\mathsf{key}, \mathsf{value}>$, where $\mathsf{key}$ and $\mathsf{value}$ are both 32-bit integers. We use band join to test the performance of PanJoin processing a non-equi-join. The band join for Stream S is defined as:
$$
\bm{\mathsf{WHERE}}\; \mathsf{s.value}\; \bm{\mathsf{BETWEEN}}\; \mathsf{r.value} - \epsilon\; \bm{\mathsf{AND}}\; \mathsf{r.value} + \epsilon
$$
and vice versa for Stream R, where $\epsilon$ is used to control the selectivity. 
The throughput is mainly influenced by the following four factors: subwindow size $N_{Sub}$, batch size $N_{Bat}$, partition count $P$, partition size $N_P = N_{Sub}/P$, and selectivity $S$ which is defined as the average number of matching tuples in a full subwindow per probing tuple. We use count-based windows to present the throughput because it is easier to show the correlation between throughput and the factors it relies on, while changing the input rate of time-based windows will change both the window size and the throughput. Some data structures have their features, which are also presented in the following part of this section. RaP-Table and WiB$^+$-Tree perform insertion with 1 thread, and BI-Sort performs insertion with 8 threads. The probe operations of all data structures are paralleled with 8 threads when $N_{Bat} \geqslant 128$ and are executed with 1 thread when $N_{Bat} < 128$. 

\subsubsection{RaP-Table on CPU} \label{sec:eva_rap}

We first show the insertion throughput with a fixed subwindow size $N_{Sub}$=8M in Figure \ref{fig:rap_insert_1}. A larger batch size provides better performance because tuples in the same partition can share memory accesses to the LLAT, and fewer partitions (smaller $P$) save time during binary search on the partition table. When $N_{Bat}$=8M, RaP-Table reaches its peak throughput (approximately 100M tuples/s). In Figure \ref{fig:rap_insert_2} with a fixed $P$, we observe a slight but not significant advantage of smaller subwindow size $N_{Sub}$, which proves that the insertion throughput of RaP-Table mainly relies on the scan of the partition table and memory access to the LLAT. 

Figure \ref{fig:rap_probe_1} presents the throughput of equi-join with a selectivity $S$=1, meaning that on average, a probing tuple matches one tuple in the subwindow. Here, smaller $N_P$ provides better performance because fewer tuples are accessed per probing tuple. Additionally, a smaller subwindow runs faster because it has a smaller partition table to scan. When $S$ increases, as shown in Figure \ref{fig:rap_probe_2}, the throughput begins to decrease. Meanwhile, when $S$ is small, a larger partition count $P$ provides higher throughput since it corresponds to a smaller $N_P$. The performance of non-equi-join probing shown in Figure \ref{fig:rap_probe_3} is similar to the equi-join probing: smaller $S$ and larger $P$ provides higher throughput. 
%except that there is a throughput drop in each line, e.g., $P$=256K at selectivity $S=2^6$. Here, because $S > N_P$, the lower and upper bounds of the join condition stays in different partitions which requires extra processing time. Note that with a large selectivity, non-equi-join is faster than equi-join because in non-equi-join RaP-Table does not actually compare the values in the middle partitions but copies them straight to the result, which is faster than comparing the same amount of tuples as is done in the equi-join probing. 

In Figure \ref{fig:rap_adjust}, we also present the performance, i.e. the normalized MAE (mean average error),  under the multimodal normal distributions with legends ``N(normalized $\sigma$, modal count, $P$)'', uniform distributions with legends ``U(normalized range, modal count, $P$)''， and a real dataset of Youtube videos (first file, depth 4) \cite{chen2008youtube}. In the Youtube data, the values (view counts) follow a rank-size distribution, where 99\% fall in the 1\% of the data range or 0.01\% of the range of a 32-bit integer. Initially, the partition table assumes that the value is evenly distributed among the range of a 32-bit integer.
We can observe that subwindows with smaller $P$ require fewer iterations to converge and deliver more balanced outcomes (with a lower MAE). Under each distribution, RaP-Table is able to converge in 3 iterations, which proves our statement in Section \ref{adjust_algo}.

\subsubsection{WiB$^+$-Tree on CPU}

\begin{figure*}[!htb]
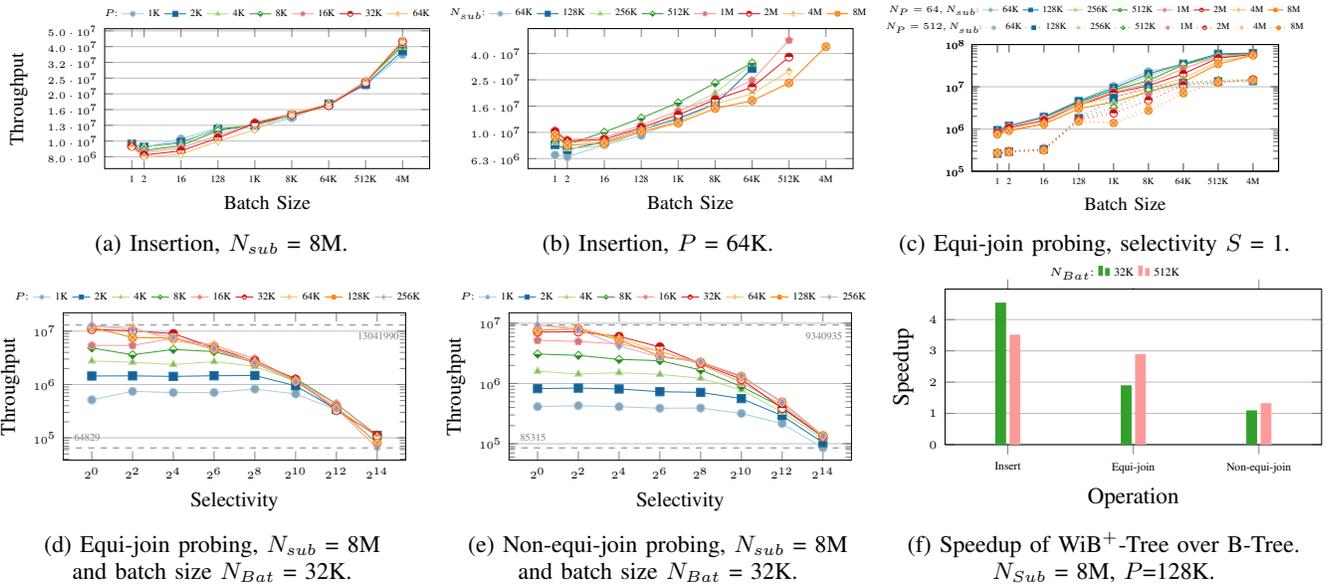

\captionsetup[subfigure]{justification=centering}
    \centering
    \begin{subfigure}[b]{0.32\textwidth}

\centering
\includegraphics[width=\textwidth]{figures/wib_insert_1.tikz}
\caption{Insertion, $N_{sub}$ = 8M.}
\label{fig:wib_insert_1}

    \end{subfigure}%
    \begin{subfigure}[b]{0.32\textwidth}

\centering
\includegraphics[width=\textwidth]{figures/wib_insert_2.tikz}
\caption{Insertion, $P$ = 64K.}
\label{fig:wib_insert_2}
    
    \end{subfigure}%
\begin{subfigure}[b]{0.32\textwidth}
\centering
\includegraphics[width=\textwidth]{figures/wib_probe_1.tikz}
\caption{Equi-join probing, selectivity $S$ = 1.}
\label{fig:wib_probe_1}
    
    \end{subfigure}
        \begin{subfigure}[b]{0.32\textwidth}
\centering
\includegraphics[width=\textwidth]{figures/wib_probe_2.tikz}
\caption{Equi-join probing, $N_{sub}$ = 8M \\ and batch size $N_{Bat}$ = 32K.}
\label{fig:wib_probe_2}

\end{subfigure}
 \begin{subfigure}[b]{0.32\textwidth}
 \includegraphics[width=\textwidth]{figures/wib_probe_3.tikz}
\caption{Non-equi-join probing, $N_{sub}$ = 8M  \\ and batch size $N_{Bat}$ = 32K.}
\label{fig:wib_probe_3}
 \end{subfigure}
 \centering
 \begin{subfigure}[b]{0.32\textwidth}
 \includegraphics[width=\textwidth]{figures/wib_comp.tikz}
\caption{Speedup of WiB$^+$-Tree over B-Tree. $N_{Sub}$ = 8M, $P$=128K.}
\label{fig:wib_comp}
 \end{subfigure}
%  \begin{subfigure}[b]{0.32\textwidth}
%  \centering
%   \includegraphics[width=\textwidth]{figures/wib_width.tikz}
%\caption{Adjustment algorthm. $N_{sub}$ = 510K for Youtube and $N_{sub}$ = 8M for the rest.}
%\label{fig:wib_width}
%   \end{subfigure}
   \caption{Performance of WiB$^+$-Tree.}
      \label{fig:wib}
\end{figure*}

We use the same metrics as for RaP-Table to test the performance of WiB$^+$-Tree. Here, the partition count $P$ is the number of leaf nodes in the tree. The number is approximate because the tree structure may vary with different randomly generated data. Similar to RaP-Table, in Figure \ref{fig:wib_insert_1} and Figure \ref{fig:wib_insert_2}, we can also observe the benefit brought by a large batch size for insertion operations.  However, with a fixed $N_{Sub}$, the insertion throughput of WiB$^+$-Tree is not sensitive to the value of $P$, as shown in Figure \ref{fig:wib_insert_1}. This is because we do not retain the tuples sorted in the leaf node; therefore the size of leaf node (=$N_{Sub} / P$) does not affect the throughput much. Figure \ref{fig:wib_insert_2} shows that with a fixed $P$, larger subwindows ($>$512K) normally perform slower because the LLAT is too large to fit in the cache, which leads to more memory accesses and lower throughput.

In Figure \ref{fig:wib_probe_1}, we can observe a similar  impact of $N_P$ on the probing throughput of WiB$^+$-Tree to RaP-Table. Figure \ref{fig:wib_probe_2} and Figure \ref{fig:wib_probe_3} also show the similar throughput decreases when the selectivity is large. Both WiB$^+$-Tree to RaP-Table can reach an ideal throughput when $P>$16K.

Figure \ref{fig:wib_comp} shows the speedup of WiB$^+$-Tree over a regular B-Tree implemented by Google \cite{btree2011google}, where $N_{Sub}$ = 8M, $P$=128K, and $S$ varies from 1 to 16K. The insertion speedup is 3.5-4.5x because WiB$^+$-Tree does not sort the leaf nodes. When the batch size $N_{Bat}$ is larger, the speedup of equi-join and non-equi-join probing becomes higher (up to 2.8x for equi-join and 1.3x for non-equi-join), which shows the efficiency of batch mode. Here, we prove the correctness of our design rationale. 

\subsubsection{BI-Sort on CPU}

\begin{figure*}[!htb]
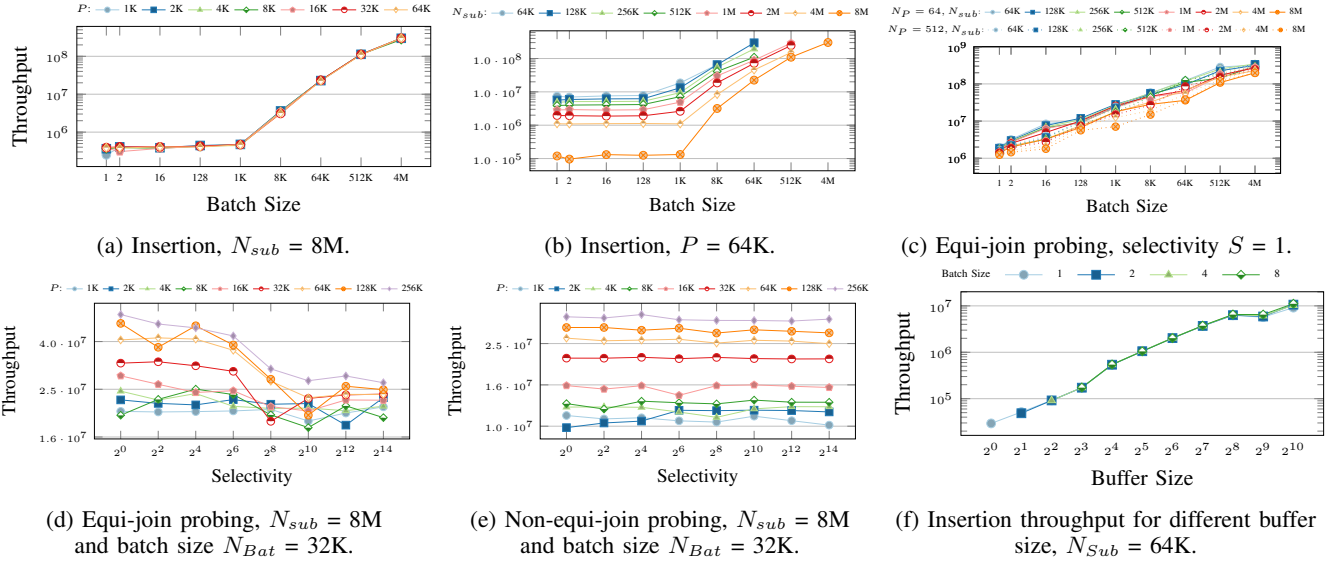

\captionsetup[subfigure]{justification=centering}
    \centering
    \begin{subfigure}[b]{0.32\textwidth}

\centering
\includegraphics[width=\textwidth]{figures/bis_insert_1.tikz}
\caption{Insertion, $N_{sub}$ = 8M.}
\label{fig:bis_insert_1}

    \end{subfigure}%
    \begin{subfigure}[b]{0.32\textwidth}

\centering
\includegraphics[width=\textwidth]{figures/bis_insert_2.tikz}
\caption{Insertion, $P$ = 64K.}
\label{fig:bis_insert_2}
    
    \end{subfigure}%
\begin{subfigure}[b]{0.32\textwidth}
\centering
\includegraphics[width=\textwidth]{figures/bis_probe_1.tikz}
\caption{Equi-join probing, selectivity $S$ = 1.}
\label{fig:bis_probe_1}
    
    \end{subfigure}
        \begin{subfigure}[b]{0.32\textwidth}
\centering
\includegraphics[width=\textwidth]{figures/bis_probe_2.tikz}
\caption{Equi-join probing, $N_{sub}$ = 8M \\ and batch size $N_{Bat}$ = 32K.}
\label{fig:bis_probe_2}

\end{subfigure}
\centering
 \begin{subfigure}[b]{0.32\textwidth}
 \includegraphics[width=\textwidth]{figures/bis_probe_3.tikz}
\caption{Non-equi-join probing, $N_{sub}$ = 8M  \\ and batch size $N_{Bat}$ = 32K.}
\label{fig:bis_probe_3}
 \end{subfigure}
  \begin{subfigure}[b]{0.32\textwidth}
  \centering
   \includegraphics[width=\textwidth]{figures/bis_buffer.tikz}
\caption{Insertion throughput for different buffer size, $N_{Sub}$ = 64K.}
\label{fig:bis_buffer}
   \end{subfigure}
   \caption{Performance of BI-Sort.}
      \label{fig:bis}
\end{figure*}

We use the same metrics as RaP-Table and  WiB$^+$-Tree to test the performance of BI-Sort. The default size of the insertion buffer is 1K. Figure \ref{fig:bis_insert_1} and Figure \ref{fig:bis_insert_2} show a significant impact of the insertion buffer: the throughput remains at the same value when $N_{Bat} < $1K. Figure \ref{fig:bis_insert_1} shows insertion of BI-Sort is not sensitive to $P$ because the index array is very small and requires little computation time compared to the main array. However, Figure \ref{fig:bis_insert_2} shows that a larger $N_{Sub}$ has a considerably slower throughput when $N_{Bat}$ is small. In these cases, main array is so large that merging the insertion buffer or tuple batch is costly. The only solution is to increase $N_{Bat}$ to amortize the merging time. 

As shown in Figure \ref{fig:bis_probe_1}, $N_P$ does not greatly affect on BI-Sort's probing throughput except when $N_{Bat}$ is small. Figure \ref{fig:bis_probe_2} and Figure \ref{fig:bis_probe_3} show the main difference and advantage of BI-Sort compared with RaP-Table and WiB$^+$-Tree: the probing throughput of BI-Sort is not sensitive to selectivity $S$ because BI-Sort only attempts to find the indices of the join result in the subwindow rather than copying the real tuples to the result. For equi-join shown in Figure \ref{fig:bis_probe_2}, as $S$ increases, each tuple may need to probe two partitions instead of one. Thus, there is a throughput decrease when $P$ is sufficiently large such that $N_P$ is smaller than $S$, whereas Figure \ref{fig:bis_probe_3} has no decreases because during non-equi-join, BI-Sort always checks the upper and lower bounds of join conditions per probing tuple. Figure \ref{fig:bis_probe_2} and \ref{fig:bis_probe_3} also tell us that a larger $P$ improves the throughput because more tuples share the same memory access to the same target partition. 

In Figure \ref{fig:bis_buffer}, we use $N_{sub}$=64K to illustrate the impact of buffer size on handling inputs with a small batch size. As shown, the throughput with a large buffer size is 2-3 orders of magnitude larger than with a small buffer size. However, because the buffer is unsorted, the tuple batch needs to perform a nest-loop-join with the buffer during the probing operation. Therefore, we do not recommend a large buffer size which can reduce the system performance.  

\subsubsection{Comparison}

\begin{figure}[!htb]
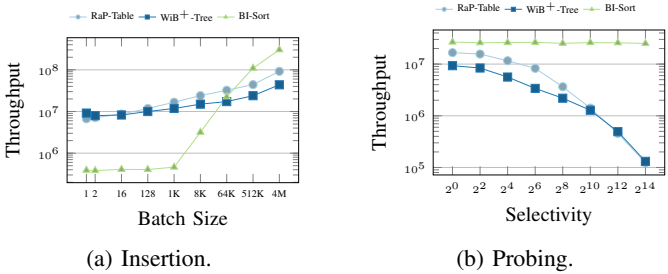

\centering
\captionsetup[subfigure]{justification=centering}
    \begin{subfigure}[t]{0.22\textwidth}
        \centering
 \includegraphics[width=\textwidth]{figures/com_insert.tikz}
\caption{Insertion.}
\label{fig:com_insert}
 \end{subfigure}\hfill%
  \begin{subfigure}[t]{0.22\textwidth}
  \centering
   \includegraphics[width=\textwidth]{figures/com_probe.tikz}
\caption{Probing.}
\label{fig:com_probe}
   \end{subfigure}
   \caption{Performance comparisons, $N_{Sub}$=8M and $P$=64K.}
   \label{fig:com}
\end{figure}

We compare the throughput of insertion and non-equi-join probing with the configuration $N_{Sub}$=8M and $P$=64K. For insertion shown in \ref{fig:com_insert} , BI-Sort outperforms the other two only when batch size is larger than 64K. When $N_{Bat} <$64K, the insertion of BI-Sort becomes the bottleneck of the system. For non-equi-join probing with $N_{Bat} =$32K shown in \ref{fig:com_probe}, when selectivity $S<$1K ($2^{10}$),  RaP-Table is 1.7-2.4x over WiB$^+$-Tree and BI-Sort is about 1.6-7x over RaP-Table. When $S$ is large, BI-Sort can be 100x over the other two because it is not sensitive to $S$, as we have stated before. Therefore, we suggest the following strategy: if $N_{Bat}$ is large or $S$ is large, we should choose BI-Sort; otherwise, if the tuple values does not vary too often and do not gradually increase, we can use RaP-Table, and if not, we should use WiB$^+$-Tree. 

\subsection{BI-Sort on FPGA}

\begin{figure}[!htb]
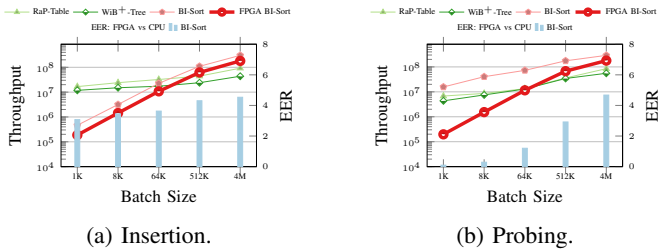

\centering
\captionsetup[subfigure]{justification=centering}
    \begin{subfigure}[t]{0.22\textwidth}
        \centering
 \includegraphics[width=\textwidth]{figures/fpga_insert.tikz}
\caption{Insertion.}
\label{fig:fpga_insert}
 \end{subfigure}\hfill%
  \begin{subfigure}[t]{0.22\textwidth}
  \centering
   \includegraphics[width=\textwidth]{figures/fpga_probe.tikz}
\caption{Probing.}
\label{fig:fpga_probe}
   \end{subfigure}
   \caption{Performance of BI-Sort on FPGA.}
   \label{fig:fpga}
\end{figure}

\begin{table}
\centering
\caption{Summary of FPGA Resources}
\label{tab:fpga}
\begin{tabular}{l | l|l|l}
\specialrule{.15em}{.05em}{.05em}
 & Used & Total & Utilization \\
 \specialrule{.15em}{.05em}{.05em}
Logic	&178647	&427200	&42\%\\
\specialrule{.025em}{.0em}{0em}
BRAM	&848	&2713	&31\% \\
\specialrule{.025em}{.0em}{0em}
DSP		& 0  	& 1,518 & 0\%\\
\specialrule{.025em}{.0em}{0em}
Clock Frequency & \multicolumn{3}{c}{252.7MHz}\\
\specialrule{.15em}{.05em}{.05em}
\end{tabular}
\end{table}

We implement our FPGA subwindow on a Terasic DE5a-Net FPGA Development Kit which contains an Arria 10 (10AX115N\-2F45E1SG) FPGA and two channels of DDR3-1066 4 GB memory. We present the performance comparison of the FPGA version with the 3 data structures on CPU in Figure \ref{fig:fpga}, where $N_{Sub}$=8M and $P$=64K. There are 8 mergers and 8 probers on the FPGA. More mergers or probers can not provide better performance because these 16 units are enough to fully utilize the memory bandwidth on FPGA. Figure \ref{fig:fpga_insert} presents the insertion throughput: when $N_{Bat}$ becomes larger than 64K, BI-Sort on FPGA is faster than RaP-Table and WiB$^+$-Tree on CPU. BI-Sort on FPGA is approximately 0.4x-0.6x over the throughput of BI-Sort on CPU because the DDR3 memory (2 channels) used by FPGA is 8.8x slower than the DDR4 memory (6 channels) used by CPU. However, the energy efficiency ratio (EER) on FPGA is approximate 4x over that on CPU, shown as bars using the right y-axis. This result occurs because the TDP (Thermal Design Power) of the CPU is 150 W and we use 8 out of 20 cores on CPU such that the power of the CPU version is 60 W, while the power of the FPGA solution is only 7.9W. For equi-join probing with $S$=1 show in Figure \ref{fig:fpga_probe},  when $N_{Bat}$ is larger than 64K, BI-Sort on FPGA provides better throughput than RaP-Table and WiB$^+$-Tree on CPU. It also provides an EER larger than 1 over BI-Sort on CPU. Therefore, when $N_{Bat}$ is large enough, BI-Sort on FPGA becomes an excellent choice when both throughput and power are concidered. Table \ref{tab:fpga} shows that our FPGA solution uses 42\% of the logic resources and 31\% of the internal block memory (BRAMs), which means that we can puts more processing units for other operations.

\subsection{System Performance}

\begin{figure*}[!htb]
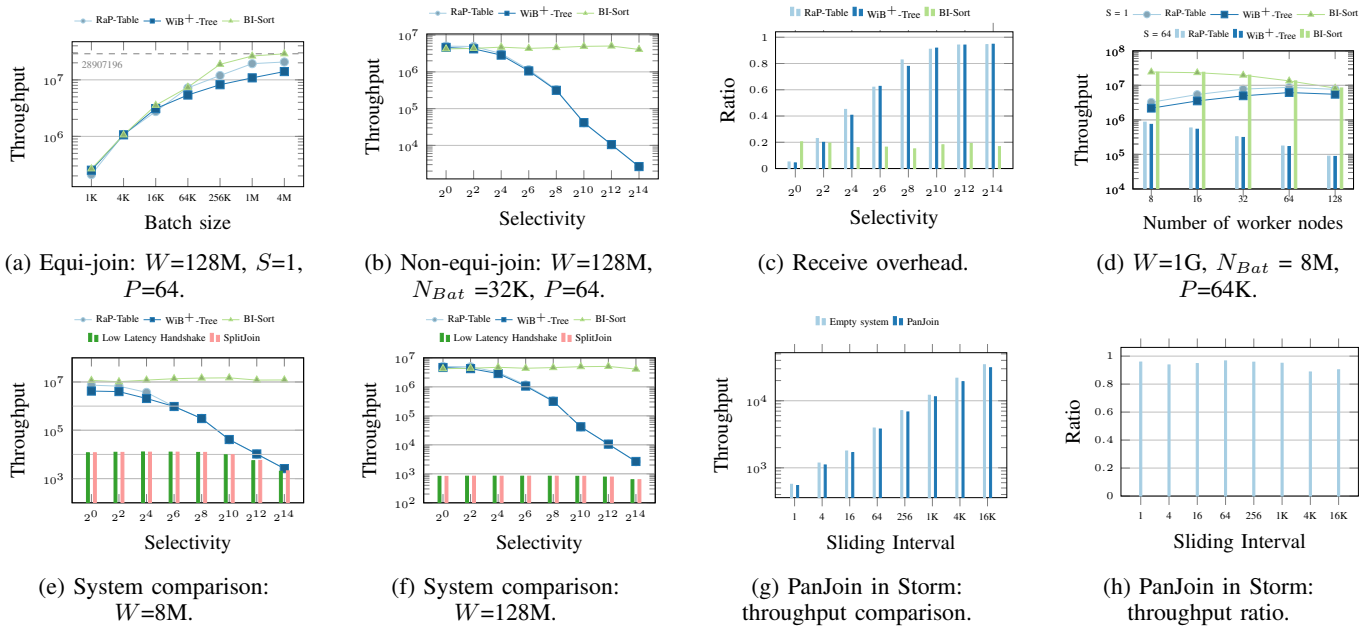

\captionsetup[subfigure]{justification=centering}
    \centering
    \begin{subfigure}[t]{0.22\textwidth}

\centering
\includegraphics[width=\textwidth]{figures/sys_per_1.tikz}
\caption{Equi-join: $W$=128M, $S$=1, $P$=64.}
\label{fig:sys_per_1}

    \end{subfigure}\hfill%
    \begin{subfigure}[t]{0.22\textwidth}

\centering
\includegraphics[width=\textwidth]{figures/sys_per_2.tikz}
\caption{Non-equi-join: $W$=128M, $N_{Bat}$ =32K, $P$=64.}
\label{fig:sys_per_2}
    
    \end{subfigure}\hfill%
\begin{subfigure}[t]{0.22\textwidth}
\centering
\includegraphics[width=\textwidth]{figures/sys_per_3.tikz}
\caption{Receive overhead. }
\label{fig:sys_per_3}
    
    \end{subfigure}\hfill
        \begin{subfigure}[t]{0.22\textwidth}
\centering
\includegraphics[width=\textwidth]{figures/sys_per_4.tikz}
\caption{$W$=1G, $N_{Bat}$ = 8M, $P$=64K.}
\label{fig:sys_per_4}

\end{subfigure}

    \begin{subfigure}[b]{0.22\textwidth}
        \centering
 \includegraphics[width=\textwidth]{figures/sys_com_1.tikz}
\caption{System comparison: $W$=8M.}
\label{fig:sys_com_1}
 \end{subfigure}\hfill%
  \begin{subfigure}[b]{0.22\textwidth}
  \centering
   \includegraphics[width=\textwidth]{figures/sys_com_2.tikz}
\caption{System comparison: $W$=128M.}
\label{fig:sys_com_2}
   \end{subfigure}\hfill%
       \begin{subfigure}[b]{0.22\textwidth}
        \centering
 \includegraphics[width=\textwidth]{figures/sys_storm_1.tikz}
\caption{PanJoin in Storm: throughput comparison.}
\label{fig:sys_storm_1}
 \end{subfigure}\hfill%
  \begin{subfigure}[b]{0.22\textwidth}
  \centering
   \includegraphics[width=\textwidth]{figures/sys_storm_2.tikz}
\caption{PanJoin in Storm: throughput ratio.}
\label{fig:sys_storm_2}
   \end{subfigure}
   \label{fig:sys}
   \caption{System performance of PanJoin.}
\end{figure*}

We test the complete PanJoin on a cluster. Each node has the same configuration as discussed in Section \ref{sec:ana_eva}. Every two nodes are connected with InfiniBand with a bandwidth of 100 Gb/s (12.5 GB/s). We first compare the throughput of equi-join with selectivity $S$=1, as shown in Figure \ref{fig:sys_per_1}. The window size $W$=128M for both streams. Each stream has 16 worker nodes and there is one subwindow per node, i.e., $N_{Sub}$=128M$/16$=8M. We find that PanJoin with BI-Sort has the highest throughput up to 28.9M/s. At this input rate, we can process 1G input tuples within 34.6 seconds. We also observe that RaP-Table is faster than WiB$^+$-Tree by approximate 1.5x when $N_{Bat}>$64K. Figure \ref{fig:sys_per_1} also suggests that if the input rate is high, we should use a large batch size such that the system can handle the data. Figure \ref{fig:sys_per_2} shows the system performance of non-equi-join with a smaller batch size ($N_{Bat}$=32K). We can observe the throughput decreases on RaP-Table and WiB$^+$-Tree when the selectivity $S$ becomes larger. Note that the throughput of RaP-Table and WiB$^+$-Tree are nearly identical because under this configuration the overhead of data transmission between manger and workers is more than 60\% of the processing time, which amortizes the throughput differences inside subwindows. Therefore, their lines are overlapped in the figure. Figure \ref{fig:sys_per_3} shows the overhead of data transmission from workers to the manager (not manager to workers). When the selectivity becomes larger than $2^{10}$ (1K), the result receiving overhead for RaP-Table and WiB$^+$-Tree can be more than 90\% of the processing time. For this reason, we marked result receiving as an optional operation in Step 5 of the system architecture (Section \ref{sec:archtect}). We also evaluate non-equi-join with large window $W$=1G processed by different numbers of worker nodes, as shown in Figure \ref{fig:sys_per_4}, where we use lines for selectivity $S$=1 and bars for $S$=64. Here, adding more worker nodes does not always gives better performance because the network communication overhead is dominating the execution time. Still, the system with BI-Sort provide a throughput of approximately 10M/s ($10^7$), where the other two data structures are 1-10x slower when $S$=1 and 10-100x slower when $S$=64.% and are sensitive to selectivity.

We compare PanJoin with Low Latency Handshake Join and SplitJoin (ScaleJoin), which are adapted to support large $N_{Bat}=$32K and more subwindows/nodes. We consider SplitJoin and ScaleJoin together because their architectures are similar. With window size $W$=8M and 16 worker nodes in Figure \ref{fig:sys_com_1}, the speedup of BI-Sort is more than 1000x over Low Latency Handshake Join and SplitJoin (ScaleJoin), while RaP-Table and WiB$^+$-Tree is more than 100x when the selectivity is small, and RaP-Table is approximate 1.5x over WiB$^+$-Tree. When the selectivity is larger than 4K, RaP-Table and WiB$^+$-Tree have the same throughput as Handshake Join and SplitJoin (ScaleJoin). When $W$ increases to 128 in Figure \ref{fig:sys_com_2}, the speedup of BI-Sort increases to approximately 5000x, while the RaP-Table's and WiB$^+$-Tree's speedup are approximately 100x-5000x when $S\leqslant$1K and 2-100x when $S>$1K. In this way, we show how powerful PanJoin is as an integrated design compared with the existing stream join solutions, which use a nested-loop join inside their subwindows/nodes.

We also attempt to integrate PanJoin into Apache Storm \cite{toshniwal2014storm}. Because the system overhead is too large compared with the pure PanJoin solution, we compare the performance of integrated PanJoin with a system where every subwindow is empty, i.e., the join processor does nothing but receive the input tuples and discard them afterward. Figure \ref{fig:sys_storm_1} shows the absolute throughput and Figure \ref{fig:sys_storm_2} shows the throughput ratio of PanJoin to the empty system, where we set $N_{Sub}$=1M. Here, throughput ratio is always approximately 90 percent, which proves that PanJoin works well with Storm and is fast enough for an existing stream processing engine.

\section{Conclusions} \label{sec:con}

In this paper, we present a stream join solution called PanJoin that has high throughput, supports non-equi-join, and adapts to skewed data. We present the three new data structures for subwindows to manage their data and provide a strategy that users can choose. Our evaluation proved that the performance of PanJoin is more than three orders of magnitude higher than several recently proposed stream join solutions. The limitation of PanJoin on large window sizes is the network bandwidth. Since InfiniBand is one of the fastest solutions on the market, we expect that the future network technology will solve this limitation.

\bibliographystyle{ieeetr}
\bibliography{sample-bibliography} 

\end{document}